\input amstex
\documentstyle{amsppt}


\font\eightrm=cmr8
\font\eighti=cmmi8
\font\eightsy=cmsy8
\font\eightbf=cmbx8
\font\eighttt=cmtt8
\font\eightit=cmti8
\font\eightsl=cmsl8
\font\sixrm=cmr6
\font\sixi=cmmi6
\font\sixsy=cmsy6
\font\sixbf=cmbx6

\catcode`@=11
\newskip\ttglue

\def\eightpoint{\def\rm{\fam0\eightrm}%
  \textfont0=\eightrm \scriptfont0=\sixrm \scriptscriptfont0=\fiverm
  \textfont1=\eighti  \scriptfont1=\sixi  \scriptscriptfont1=\fivei
  \textfont2=\eightsy \scriptfont2=\sixsy \scriptscriptfont2=\fivesy
  \textfont3=\tenex   \scriptfont3=\tenex \scriptscriptfont3=\tenex
  \textfont\itfam=\eightit \def\it{\fam\itfam\eightit}%
  \textfont\slfam=\eightsl \def\sl{\fam\slfam\eightsl}%
  \textfont\ttfam=\eighttt \def\tt{\fam\ttfam\eighttt}%
  \textfont\bffam=\eightbf \scriptfont\bffam=\sixbf
   \scriptscriptfont\bffam=\fivebf \def\bf{\fam\bffam\eightbf}%
  \tt \ttglue=.5em plus.25em minus.15em
  \normalbaselineskip=9pt
  \setbox\strutbox=\hbox{\vrule height7pt depth2pt width0pt}%
  \let\sc=\sixrm \let\big=\eightbig \normalbaselines\rm}

\def\tenbig#1{{\hbox{$\left#1\vbox to8.5pt{}\right.\n@space$}}}
\def\ninebig#1{{\hbox{$\textfont0=\tenrm\textfont2=\tensy
  \left#1\vbox to7.25pt{}\right.\n@space$}}}
\def\eightbig#1{{\hbox{$\textfont0=\ninerm\textfont2=\ninesy
  \left#1\vbox to6.5pt{}\right.\n@space$}}}


\font\bigbf=cmbx10 scaled\magstep1
\font\smc=cmcsc10
\font\got=eufm10
\def\var{\varepsilon}
\nologo
\NoBlackBoxes

\def\sqr#1#2{{\vcenter{\vbox{\hrule height .#2pt
    \hbox{\vrule width .#2pt height#1pt \kern#1pt
    \vrule width.#2pt}
    \hrule height.#2pt}}}}

\noindent{\bigbf The Quantum Structure of Spacetime at the Planck}\smallskip
\noindent{\bigbf Scale and Quantum Fields}\bigskip
\noindent{\smc Sergio Doplicher$^1$, Klaus Fredenhagen$^2$, John E. Roberts$^3$}\bigskip
\noindent{\eightpoint $^1$ Dipartimento di Matematica, Universit\`a di Roma ``La Sapienza'', I--00185 Roma,
Italy}\smallskip
\noindent{\eightpoint$^2$ II Institut f\"ur Theoretische Physik der Universit\"at Hamburg, D--22761 Hamburg,
Germany}\smallskip
\noindent{\eightpoint$^3$ Dipartimento di Matematica, Universit\`a di Roma ``Tor Vergata'', I--00133 Roma,
Italy}\smallskip
\noindent{\eightpoint$^{1,3}$ Research supported by MRST and CNR--GNAFA}\bigskip

\noindent Commun.\ Math.\ Phys.\ {\bf 172}, 187--220 (1995)\bigskip

\noindent Received: 22 June 1994\bigskip\bigskip

{\narrower\smallskip\eightpoint\baselineskip10pt\noindent{\bf Abstract}. We propose uncertainty relations for the 
different coordinates of spacetime events, motivated by Heisenberg's principle and by Einstein's theory of 
classical gravity. A model of Quantum Spacetime is then discussed where the commutation relations exactly implement
our uncertainty relations.\par
\indent We outline the definition of free fields and interactions over QST and take the first steps to  adapting
the usual  perturbation theory. The quantum nature of the underlying spacetime replaces a local interaction by a
specific nonlocal effective interaction in the ordinary Minkowski space. A detailed study of interacting QFT and of
the smoothing of ultraviolet divergences is deferred to a subsequent paper.\par
\indent In the classical  limit where the Planck length goes to zero, our Quantum Spacetime reduces to  the
ordinary  Minkowski space times a two component space whose components are homeomorphic to the tangent bundle
$TS^2$ of the 2--sphere. The relations with Connes' theory of the standard model will be studied
elsewhere.\smallskip}
\vskip3.5truecm

\noindent{\bf 1. Introduction}\medskip

\noindent
It is generally believed that the picture of spacetime as a manifold $M$ locally modelled on the flat Minkowski 
space $M_0=\Bbb R^4$ should break down at very short distances of the order of the Planck length
$$\lambda_P=\left({G\hbar\over c^3}\right)^{1/2}\simeq1.6\times10^{-33}\text{ cm.}$$
Limitations in the possible accuracy of localization of spacetime events should in fact be a feature of a Quantum 
Theory incorporating gravitation.

There have been investigations on possible mechanisms leading to such limitations in the context of string  theory
[1,2], in Ashtekar's approach to quantum gravity [3], and, in a more formal way, in the context of Quantum Groups
[4,5]. These different approaches have led to different limitations and, more significantly, to different pictures
of spacetime where gravitational effects in the small are necessarily strong (spacetime foam [6]).

Our proposal differs radically: attempts to localize with extreme precision cause gravitational collapse  so that
spacetime below the Planck scale has no operational meaning. We elaborate on this well known remark and are led to
spacetime uncertainty relations. In our proposal, spacetime has a quantum structure intrinsically implying those
relations. Thus the impossibility of giving an operational meaning to spacetime in the small is incorporated in the
mathematical structure of the model.

Similar models can be found in the work of J. Madore [7], where, however, no attempt was made to motivate  the
algebraic structure by an operational analysis of localization.

We thus propose that spacetime ought to be described as a {\it non--commutative manifold\/}, i.e.  the commutative
algebra ${\Cal C}_0(M)$ of complex continuous functions on $M$ vanishing at infinity should be replaced by a
non--commutative algebra ${\Cal E}$, and points of $M$ by pure states on ${\Cal E}$. The aim of this paper is to
propose an algebra ${\Cal E}$ describing Quantum Spacetime and to look for a formulation of QFT over QST.

We now formulate some criteria for the choice of ${\Cal E}$. We are interested in elementary particle  physics,
that is in describing idealized situations where only few colliding particles are present. Therefore, as a first
state of our project, we are interested in a non--commutative variant of the flat spacetime $M_0$, deviating from
$M_0$ only at very short distances, which fulfils the following principles:
{\bigskip} 
{\it
\item{1)} The commutation relations in ${\Cal E}$ should be motivated by operationally meaningful  uncertainty
relations between the different coordinates of spacetime events.
\item{2)} The flat spacetime $M_0$ should appear (possibly as a factor) in the large scale limit of ${\Cal E}$.
\item{3)} The full Poincar\'e group should act as symmetries on ${\Cal E}$\/}.
{\bigskip}
Condition 3 is motivated and made possible by 2) and allows us to adopt Wigner's description of  elementary
particles in terms of irreducible representations of the Poincar\'e group. Parity or time reversal symmetry
breaking might be features of specific interactions but the quantum spacetime should be, as ordinary spacetime,
reflection symmetric. Therefore, in 3) the full Poincar\'e group is required to yield symmetries of the quantum
spacetime.

In Sect. 2 we explore the limitations of localization measurements which are due to the possible  creation of black
holes by concentration of energy.

We find the uncertainty relations
$$\Delta x_0(\Delta x_1+\Delta x_2+\Delta x_3)\gtrsim\lambda^2_P\ ,$$
$$\Delta x_1\Delta x_2+\Delta x_2\Delta x_3+\Delta x_3\Delta x_1\gtrsim
\lambda^2_P\ ,$$
which are implied by those limitations but do not necessarily imply them. In Sect. 3 we find  algebraic relations
which imply these uncertainty relations. They have the form
$$\eqalign{
q_\mu&=q^*_\mu\ ,\cr
[[q_\mu,q_\nu],q_\rho]&=0\ ,\cr
[q_\mu,q_\nu][q^\mu,q^\nu]&=0\ ,\cr
\left({1\over8}\,[q_\mu,q_\nu][q_\rho,q_\sigma]\varepsilon^{\mu\nu\rho\sigma}\right)^2&=
\lambda^8_P\ .\cr}$$
The resulting algebra has a centre generated by the commutators $iQ_{\mu\nu}=[q_\mu,q_\nu]$.  The physical meaning
of this centre has still to be understood. It is responsible for the fact that in the large scale limit, performed
at fixed spectral values of the centre, the resulting classical space is $M_0\times\Sigma$, where
$$\eqalign{
\Sigma&=\{(\sigma_{\mu\nu}),\sigma_{\mu\nu}=-\sigma_{\nu\mu},\sigma_{\mu\nu}\sigma^{\mu\nu}=0\ ,
\ (1/8\sigma_{\mu\nu}\sigma_{\rho\sigma}\varepsilon^{\mu\nu\rho\sigma})^2=1\}\cr
&\simeq TS^2\times\{1,-1\}\ .\cr}$$
In Sect. 4 we define a $C^*$--algebra ${\Cal E}$ to which the operators $q_\mu$ are affiliated.  ${\Cal E}$ turns
out to be isomorphic to the algebra $C_0(\Sigma,{\Cal K})$ of continuous functions from $\Sigma$ to ${\Cal K}$
vanishing at infinity, where ${\Cal K}$ is the algebra of compact operators in a fixed separable Hilbert space.
States on ${\Cal E}$ describe the possible localization of events. For optimal localization in a specific Lorentz
frame, in the sense that $\Sigma(\Delta q_\mu)^2$ is minimal, the state must be concentrated on spectral values of
the centre in a compact submanifold $\Sigma^{(1)}$ of $\Sigma$ with $\Sigma^{(1)}\simeq S^2\times\{1,-1\}$.

In Sect. 5 we develop calculus on ${\Cal E}$. In particular we define a spacetime integral as a positive 
Poincar\'e invariant trace and integrals over spatial hyperplanes as positive weights. The existence of these
latter integrals will be crucial for introducing an interaction and depends on the fact that the uncertainty
relations admit an absolutely precise determination  of time at the cost of complete uncertainty in at least one
spatial coordinate.

In Sect. 6 we take the first steps towards quantum field theory on the quantum spacetime.  We define free fields
and show that their commutator at spacelike distances decreases like a Gaussian. We give a formal recipe for
defining interaction Hamiltonians, interacting fields and the perturbative expansion of the $S$--matrix. This
expansion could be derived from a specific nonlocal effective interaction on ordinary Minkowski space where the
nonlocal corrections are at least quadratic in $\lambda_P$.

Gauge theories on the quantum spacetime should be formulated in the framework of non--commutative  geometry [8].
More substantial deviations from theories on classical spacetime are to be expected; quantum electrodynamics, for
example, will be a non--Abelian gauge theory. The occurrence of the two point set $\{1,-1\}$ in the classical limit
recalls Connes' theory of the standard model [8,9]. We hope to take up these problems elsewhere.

The quantum aspects of gravitation might, however, well lead to a more drastic deviation from  the classical
structure of spacetime than is shown in our model, which is motivated by semiclassical arguments pertaining to
classical gravity.

Some of the structures discussed in this paper have already appeared in the literature in different  contexts. Thus
our commutation relations occur in the theory of charged particles is constant electromagnetic fields (see e.g.
[10]). A Euclidean version of fields on a non--commutative spacetime related to ours was proposed in [11]. The
Schwinger model on Madore's fuzzy sphere [7] was discussed in [12]. For other models of quantum spacetime see e.g.
[13].

A less technical version of our main results has appeared in [14].

\beginsection{2. Spacetime Uncertainties}

We start our discussion by pointing out that combining Heisenberg's uncertainty principle with  Einstein's theory
of classical gravity leads to the conclusion that ordinary spacetime loses any operational meaning in the small.

Measuring a spacetime coordinate with great accuracy $a$ causes an uncertainty in momentum of  the order ${1\over
a}$ (unless otherwise stated, we will use natural units $\hbar=c=G=1$). Neglecting rest masses, an energy of the
order ${1\over a}$ is transmitted to the system and concentrated at some time in the localization region. The
associated energy--momentum tensor $T_{\mu\nu}$ generates a gravitational field which, in principle, should be
determined by solving Einstein's equations for the metric $\eta_{\mu\nu}$,
$$R_{\mu\nu}-{1\over2}\,R\eta_{\mu\nu}=8\pi T_{\mu\nu}\ .\eqno(2.1)$$
The smaller the uncertainties $\Delta x_\mu$ in the measurement of coordinates,  the stronger will be the
gravitational field generated by the measurement. When this field becomes so strong as to prevent light or other
signals from leaving the region in question, an operational meaning can no longer be attached to the localization.
 
Our task is now to investigate how localization is restricted by requiring that no black hole  is produced in the
course of measurement. Since pair creation is important in processes involving high energy transfer, the framework
of quantum field theory has to be used.

Our information on the localization of events in spacetime is obtained by using operations  which prepare a state
localized in a region with sides of lengths $\Delta x_\mu$, that we take as a measure of the uncertainties in the
spacetime coordinates of an event. We will admit only those states whose associated energy--momentum tensor, taken
as a source in Einstein's equation, does not generate closed trapped surfaces in the sense of
Penrose\footnote{Cf. [15, 8.2]: we are grateful to D. Boccaletti for calling our attention to this reference.}.

As a consequence, the $\Delta x_\mu$ will be subject to some restrictions preventing them  from being
simultaneously arbitrarily small. We may then pose our\medskip

\noindent{\bf Uncertainty Problem}: {\it Find the restrictions on the uncertainties  $\Delta x_0,\dots,\Delta x_3$
valid in all admissible states\/}.\medskip

To formulate our problem precisely, we turn to the free neutral scalar field and use the  coherent states as a
model for a class of states prepared by such operations. The corresponding state vector have the form
$$\Phi=e^{i\varphi(f)}\Omega\ ,\eqno(2.2)$$
where $\Omega$ denotes the vacuum state vector and $f$ a real smooth test function with compact  support where the
state (2.2) is strictly localized [16,17]. We have to choose the function $f$ so that $\Phi$ differs significantly
from the vacuum in the support of $f$, a region whose extent is characterized by $\Delta x_\mu$, $\mu=0,\dots,3$.
To this end we require the state to be significantly different from its translate through $\Delta x_\mu$. This may
be expressed by the condition
$$|(\Phi,P_\mu\Phi)|\geq(\Delta x_\mu)^{-1}\ ,$$
which will be analyzed more thoroughly elsewhere.

The mean energy--momentum tensor of the state induced by (2.2) is given by
$$t_{\mu\nu}(x;f)=(\Phi,:T_{\mu\nu}(x):\Phi)\ ,\eqno(2.3)$$
where $:\ T_{\mu\nu}\ :$ is the {\it normal form\/} of the energy--momentum density of the field
$$T_{\mu\nu}(x)={\partial\varphi\over\partial x^\mu}\,(x){\partial\varphi\over\partial x^\nu}\,
(x)+{1\over2}\,\left(m^2\varphi(x)^2-{\partial\varphi\over\partial x^\lambda}\,(x)
{\partial\varphi\over\partial x_\lambda}\,(x)\right)g_{\mu\nu}\ ,\eqno(2.4)$$
$$:\ T_{\mu\nu}(x):=T_{\mu\nu}(x)-(\Omega,T_{\mu\nu}(x)\Omega)\ .\eqno(2.5)$$
The mean energy--momentum tensor (2.3) coincides with the energy--momentum density associated  with a suitable
solution $\psi_f$ of the Klein--Gordon equation, $(\square+m^2)\psi_f=0$, using the expression (2.4) in classical
field theory. Here
$$\eqalignno{
\psi_f(x)&=Im\int_{\Omega^+_m}e^{-ikx}\hat f(k)d\Omega_m(k)\cr
&=\int\Delta_m(x-y)f(y)d^4y&(2.6)\cr}$$
(where $d\Omega_m$ is the invariant measure on the positive energy mass $m$ hyperboloid  $\Omega^+_m$ and
$\Delta_m$ the commutator function for the free scalar field of mass $m$). By the support properties of $\Delta_m$,
$\psi_f$ is localized at some time with the same accuracy $\Delta x_1$, $\Delta x_2$, $\Delta x_3$ as $\Phi$;
furthermore, by (2.6), $\psi_f$ cannot be a positive energy solution since $f$ has compact support. However,
$\psi_f$ has the same mean energy
$$(\psi_f,h\psi_f)=(\Phi,H\Phi)\ ,\eqno(2.7)$$
where $h$ is the one particle Hamiltonian and $H$ the free field Hamiltonian.  Thus our problem is equivalent to
its variant dealing only with wave functions. We will not try to solve this problem here; instead we propose as an
ansatz the following {\it spacetime uncertainty relations\/} in generic units:
$$\Delta x_0\sum^3_{j=1}\Delta x_j\gtrsim\lambda^2_P\,\eqno(2.8)$$
$$\sum^3_{j<k=1}\Delta x_j\Delta x_k\gtrsim\lambda^2_P\ .\eqno(2.9)$$
We will motivate these relations heuristically limiting ourselves to a crude estimate,  where (2.1) is replaced by
the linearized equations, the components of $T_{\mu\nu}$ with $(\mu,\nu)\not=(0,0)$ are neglected, and the total
energy $E$, where $E\sim{1\over\Delta x_0}$ if $\Delta x_0$ is very small, is supposed to be distributed with
constant density $\rho(-t)$ at negative times $x_0=-t$ over a volume with sides $\Delta x_1+t$, $\Delta x_2+t$,
$\Delta x_3+t$.

In other words we assume uniform spreading with all speeds not exceeding the speed of  light and localization
around the origin, up to $\Delta x_1$, $\Delta x_2$, $\Delta x_3$, at $x_0=0$.

The gravitational potential $\varphi$ at $x_\mu\cong0$ can be evaluated as the retarded potential
$$\varphi\sim-\int{\varphi(\vec y,-r)\over r}\,d^3\vec y\sim-{1\over\Delta x_0}\,\int^\infty_0
{1\over r}\,{r^2dr\over(\Delta x_1+r)(\Delta x_2+r)(\Delta x_3+r)}\ ,\eqno(2.10)$$
and we impose the condition that photons of energy $\varepsilon$ should not be trapped, i.e. 
$\varepsilon+\varepsilon\varphi\gtrsim0$ or $-\varphi\lesssim1$.

It is easy to compute the leading behaviour of (2.10) in the three regimes:  $\Delta x_1\sim\Delta x_2\sim\Delta
x_3$; $\Delta x_1\sim\Delta x_2\gg\Delta x_3$; $\Delta x_1\gg\Delta x_2\sim\Delta x_3$. One finds $\Delta
x_0\cdot\Delta x_1\gtrsim1$ in the first two cases, and $\Delta x_0\cdot\Delta
x_1\gtrsim\ln{\Delta x_1\over\Delta x_2}$ in the third. If $\Delta x_1/\Delta x_2$ is of the order
of unity we are back to the first case, so that $\Delta x_0\cdot\Delta x_1\gtrsim1$ is the absolute
limitation, according to (2.8), in the third case too. If $\Delta x_j\ll\Delta x_0$ we must take
$E\sim{1\over\Delta x_j}$ and we find (2.9).

If $\Delta x_0$ is very large compared with $\Delta x_j$, $j=1,2,3$, we can take as an extreme  idealization a
static solution and the Schwarzschild and Kerr solutions motivate (2.9). For, if $\Delta x_1\sim\Delta
x_2\sim\Delta x_3\sim a$, we could take a spherically symmetric solution with mass ${1\over a}$, and our condition
says that $a$ should be not smaller than the Schwarzschild radius $\sim{1\over a}$, i.e. $a\gtrsim1$.

If, say $\Delta x_1\sim\Delta x_2\sim r\gg\Delta x_3\sim a$, we may take the axially symmetric  Kerr solution with
mass $M\sim{1\over a}$ and angular momentum $L$; the limit case is $L/M\sim M$ and $L\lesssim Mr$
[18, Ch. 7] which gives
$$a\cdot r\gtrsim1$$
in accordance with (2.9). We might further argue that, in the extreme situation considered,  the energy density is
actually concentrated on a thin ring region of radius $r$ and thickness $a$, so that the requirement expressed by
(2.9) that {\it at least one\/} space uncertainty is large is not in contradiction with the Kerr solution.

However condition (2.9) is actually weaker since it allows $\Delta x_1\sim a$ to be arbitrarily  small and $\Delta
x_2\sim\Delta x_3\sim r$ to be of order one.

In the next section we discuss covariant commutation relations which do imply that the uncertainty  relations (2.8)
(2.9) hold in each state over the associated algebra. We interpret these states as describing possible
localizations of measurements. 

The separation of the spacetime localization from the measurement of a local observable is due to our  classical
treatment of the gravitational effects of localization. The present approach ought to be considered as a
semiclassical approximation to a theory, presently unknown, where gravity and quantum physics are truly unified.

\beginsection{3. Quantum Conditions on Minkowski Space}

Let $A_1$, $A_2,\dots,A_n$ be elements of a complex algebra; their {\it non--commutativity\/}  can be measured by
the quantity
$$\eqalignno{
[A_1,\dots,A_n]&\equiv\sum\varepsilon_{i_1\dots i_n}A_{i_1}\dots A_{i_n}\cr
&=\det\pmatrix
A_1\dots A_n\cr
A_1\dots A_n\cr
\dots\cr
A_1\dots A_n\endpmatrix\ .&(3.1)\cr}$$
If $n=2$, this is just the commutator; if $n=3$ and for $(j,l,k)=(1,2,3)$ and cyclic permutations we set
$$[A_j,A_l]=iC_k\ ,$$
then
$$[A_1,A_2,A_3]=i\sum_kA_kC_k\ .\eqno(3.2)$$
If $n=4$, we think of operators $q_0,\dots,q_3$ describing the coordinates in Quantum   Spacetime and forming a
Lorentz vector and we will use covariant notation. Define
$$iQ_{\mu\nu}=[q_\mu,q_\nu]\ ,\quad\mu,\nu=0,1,2,3\ .\eqno(3.3)$$
Antisymmetry of $\varepsilon$ then gives
$$[q_0,q_1,q_2,q_3]=\var^{\mu\nu\lambda\rho}q_\mu q_\nu q_\lambda q_\rho=-{1\over4}\,
\var^{\mu\nu\lambda\rho}Q_{\mu\nu}Q_{\lambda\rho}\ ,$$ so that, setting
$$(*Q)^{\mu\nu}={1\over2}\,\var^{\mu\nu\lambda\rho}Q_{\lambda\rho}$$
as usual, we get
$$[q_0,\dots,q_3]=-{1\over2}\,Q_{\mu\nu}(*Q)^{\mu\nu}\ .\eqno(3.4)$$
If we denote by $\vec e$, $\vec m$ the ``electric'' and ``magnetic'' components of $Q$, respectively, i.e.
$$\eqalignno{
Q_{0j}&=e_j=(*Q)^{lk}\cr
&\qquad\qquad\qquad\qquad(j,l,k)=(1,2,3)\hbox{ or cyclic,}\cr
Q_{lk}&=m_j=(*Q)^{0j}&(3.5)\cr}$$
the two independent Lorentz invariants which can be constructed with the tensor $Q_{\mu\nu}$ are given by
$${1\over2}\,Q_{\mu\nu}Q^{\mu\nu}=\vec m^2-\vec e{\,}^2\ ,\eqno(3.6)$$
$${1\over2}\,Q_{\mu\nu}(*Q)^{\mu\nu}=-[q_0,\dots,q_3]=\vec e\cdot\vec m+\vec m\cdot\vec e\ .\eqno(3.7)$$
These expressions are invariant under Poincar\'e transformations
$$q\to\Lambda q+a\cdot I\ ,$$
$$a\in\Bbb R^4\ ,\qquad\Lambda\in L^\uparrow_+\ ;\eqno(3.8)$$
and total reflections $q\to-q$, but (3.7) is {\it not\/} separately invariant under space or  time reflections
since the sign of (3.7) changes. Therefore, the natural conditions which are Lorentz invariant and symmetric in
$\vec e$ and $\vec m$ are the following:
$$Q_{\mu\nu}Q^{\mu\nu}=0\ ,$$
$${1\over4}\,[q_0,q_1,q_2,q_3]^2=I\ .\eqno(3.9)$$
From now on, $q_0$, $q_1$, $q_2$, $q_3$ will be always assumed to be selfadjoint operators  acting on a Hilbert
space (or affiliated to a $C^*$--algebra, cf. [19] or Appendix A), and the operators $-i[q_\mu,q_\nu]$ will be
assumed to have {\it selfadjoint closures\/} $Q_{\mu\nu}$.

We will now show that the conditions (3.9) do yield the uncertainty relations (2.8), (2.9)  {\it provided the
$Q_{\mu\nu}$ are central\/}, i.e. commute\footnote{Throughout this paper two selfadjoint operators will be said
to commute if they commute strongly, i.e. their continuous functions vanishing at infinity (or, equivalently, their
spectral resolutions) commute.} with the $q_\mu$'s
$$[q_\lambda,Q_{\mu\nu}]=0\ ;\qquad\lambda,\mu,\nu=0,\dots,3\ .\eqno(3.10)$$
We will call Eq. (3.9) and (3.10) the {\it Quantum Conditions\/}.

For each selfadjoint operator $A$ on ${\Cal H}$ and unit vector $x\in{\Cal H}$, we will  say that the vector state
$\omega=\omega_x$ is in the domain of $A$ if $x\in{\Cal D}_A$, so that
$$\omega(A^2)\equiv(Ax,Ax)<\infty\ .$$

The uncertainty $\Delta_\omega A$ is defined by $(\Delta_\omega A)^2=\omega((A-\omega(A) \cdot
I)^2)=\omega(A^2)-\omega(A)^2$ and $\omega$ is said to be {\it definite\/} on $A$ if and only if $\Delta_\omega
A=0$, i.e. $x$ is an eigenvector of $A$. The same applies if $A$ is affiliated to a $C^*$--algebra {\got A} and
$\omega\in{\Cal S}(\text{\got A})$ is in the domain of $A$ (cf. Appendix A).\bigskip

\noindent{\bf 3.1 Theorem}. {\sl Let the four selfadjoint coordinate operators $q_0,\dots,q_3$ fulfill  the Quantum
Conditions {\rm (3.9), (3.10)}. For each state $\omega$ in the domain of the $[q_\mu,q_\nu]$, we have
$$\Delta_\omega q_0\sum^3_{j=1}\Delta_\omega q_j\geq{1\over2}\ ,\eqno(3.11)$$
$$\sum_{1\leq j<k\leq3}\Delta_\omega q_j\Delta_\omega q_k\geq{1\over2}\ .\eqno(3.12)$$}

The proof will follow easily from the following propositions.\bigskip

\noindent{\bf 3.2 Proposition}. {\sl Let $q_0,\dots,q_3$ fulfill {\rm (3.9)} and $\omega$ be a state in  the domain
of the $[q_\mu,q_\nu]$'s, which is definite on $[q_0,q_1,q_2,q_3]$. Then
$$\Delta_\omega q_0\cdot\sum^3_{j=1}\Delta_\omega q_j+{1\over2}\,\sum^3_{j=1}\Delta_\omega e_j
\geq{1\over2}\ ,\eqno(3.13)$$
$$\sum_{1\leq j<k\leq3}\Delta_\omega q_j\Delta_\omega q_k+{1\over2}\,\sum^3_{j=1}\Delta_\omega m_j\geq{1\over2}\ .
\eqno(3.14)$$}\bigskip

\noindent{\bf 3.3 Proposition}. {\sl Let {\got A} be a $C^*$--algebra with unit $I$ and $\nu$ a (regular) 
probability measure on the state space ${\Cal S}(\text{\got A})$ with barycentre $\omega\in{\Cal S}(\text{\got
A})$, i.e.
$$\omega(A)=\int_{{\Cal S}(A)}\varphi(A)d\nu(\varphi)\ ,\quad A\in\text{\got A}\ .\eqno(3.15)$$
For any selfadjoint elements $A$, $B\in\text{\got A}$ we have
$$\Delta_\omega(A)\geq\int_{{\Cal S}(\text{\got A})}\Delta_\varphi(A)d\nu(\varphi)\ ,\eqno(3.16)$$
$$\Delta_\omega(A)\cdot\Delta_\omega(B)\geq\int_{{\Cal S}(\text{\got A})}\Delta_\varphi(A)
\Delta_\varphi(B)d\nu(\varphi)\ .\eqno(3.17)$$}\medskip

\noindent{\it Proof of Theorem 3.1\/}. Let the state $\omega$ be definite on each $Q_{\mu\nu}$;  then $\omega$ is
definite on $[q_0,\dots,q_3]$ (cf. (3.4)) and $\Delta_\omega(e_j)=\Delta_\omega(m_j)=0$, $j=1,2,3$, so that (3.11),
(3.12) follow from (3.13), (3.14) of Proposition 3.2. If $\omega$ is any state in the domain of the
$[q_\mu,q_\nu]$'s, it suffices, by Proposition 3.3 and Appendix A, to write $\omega$ as the barycentre of a
(regular) probability measure carried by states definite on the $Q_{\mu\nu}$'s.

Since the $Q_{\mu\nu}$ are central, $f(Q_{\mu\nu})$, $f\in{\Cal C}_0(\Bbb R)$, will lie in the centre of  the
$C^*$--algebra {\got A} generated by $f(Q_{\mu\nu})$ and $g(q_\mu)$, $\mu$, $\nu=0,1,2,3$, $f,g\in{\Cal C}_0(\Bbb
R)$. For each state $\omega$ on {\got A} in the domain of the $[q_\mu,q_\nu]$'s, the central decomposition of
$\omega$ [20, 4.8] will provide a (regular) measure carried by factor states, hence definite on $f(Q_{\mu\nu})$ and
in the domain of $[q_\mu,q_\nu]$ by Appendix A. $\square$\bigskip

\noindent{\it Proof of Proposition 3.2\/}. If $A$, $B$, $C$ are selfadjoint operators with $[A,B]^-=iC$,  then for
each state $\omega$ in the domain of $[A,B]$ we have
$$\omega(C^2)=\omega(C)^2+(\Delta_\omega(C))^2\leq(2\Delta_\omega(A)\cdot\Delta_\omega(B))^2+
(\Delta_\omega C)^2\ .\eqno(3.18)$$
If $\omega$ is as in the statement of the proposition we have $\omega([q_0,\dots,q_3])=\pm2$ and by (3.7),
$$\text{Re }\omega(\vec e\cdot\vec m)={1\over2}\,\omega(\vec e\cdot\vec m+\vec m\cdot\vec e)=\pm1\ ,$$
so that $|\omega(\vec e\cdot\vec m)|\geq1$. Since by the Schwarz inequality
$$\eqalign{
|\omega(\vec e\cdot\vec m)|&\leq\sum_j\omega(e^2_j)^{1/2}\omega(m^2_j)^{1/2}\cr
&\leq\left(\sum_j\omega(e^2_j)\right)^{1/2}\left(\sum_j\omega(m^2_j)\right)^{1/2}\ ,\cr}$$
we get
$$\omega(\vec e)^2\cdot\omega(\vec m^2)\geq1\ .$$
Since by (3.6), (3.9) $\omega(\vec e{\,}^2)=\omega(\vec m^2)$ we have $\omega(\vec e{\,}^2)\geq1$,  $\omega(\vec
m^2)\geq1$. If we recall definitions (3.3), (3.5) and apply (3.18), we get (3.13) and (3.14) using
$(a+b+\dots+c)^2\geq a^2+b^2+\dots+c^2$ for non--negative $a,b,\dots,c$. $\square$\medskip

\noindent{\it Proof of Proposition 3.3\/}. With $X\in\text{\got A}$, $X=X^*$, we have
$$\Delta_\omega(X)^2=\omega((X-\omega(X)I)^2)=\int\varphi((X-\omega(X)\cdot I)^2)d\nu(\varphi)\ .$$
But for each $\psi\in{\Cal S}(\text{\got A})$ and $\lambda\in\Bbb R$ we have
$$\eqalign{
\psi((X-\lambda\cdot I)^2)&=\psi((X-\psi(X)\cdot I)^2)+(\lambda-\psi(X))^2\cr
&\geq\psi((X-\psi(X)\cdot I)^2)\ ,\cr}$$
so that we also have
$$\eqalignno{
\Delta_\omega(X)^2&\geq\int\varphi((X-\varphi(X)\cdot I)^2)d\nu(\varphi)\cr
&=\int(\Delta_\varphi X)^2d\nu(\varphi)\ .&(3.19)\cr}$$
Since $\nu$ is a probability measure, the constant function equal to 1 is square summable with unit  $L^2$--norm
and by the Schwarz inequality
$$\eqalign{
\int\Delta_\varphi X d\nu(\varphi)&\leq\left(\int(\Delta_\varphi X)^2d\nu(\varphi)\right)^{1/2}\leq\Delta_\omega X\ ,\cr
\int\Delta_\varphi A\cdot\Delta_\varphi Bd\nu(\varphi)&\leq\left(\int(\Delta_\varphi A)^2
d\nu(\varphi)\right)^{1/2}\cdot\left(\int(\Delta_\varphi B)^2d\nu(\varphi)\right)^{1/2}\cr
&\leq\Delta_\omega A\cdot\Delta_\omega B\ ,\cr}$$
where we have used (3.19) repeatedly. $\square$\medskip

We now want to consider realizations of (3.9) and (3.10) through operators on Hilbert space.  As we have already
indicated, the $q_\mu$ will be supposed to be self--adjoint operators such that $-i[q_\mu,q_\nu]$ have
self--adjoint closures $Q_{\mu\nu}$ commuting strongly with $q_\lambda$ for all $\mu$, $\nu$, $\lambda$. We wish,
however, to impose a further condition namely that the commutation relations between the $q_\mu$'s can be
integrated in {\it Weyl form\/}:
$$e^{i\alpha_\mu q^\mu}e^{i\beta_\mu q^\nu}=e^{-(i/2)\alpha_\mu Q^{\mu\nu}\beta_\nu}
e^{i(\alpha+\beta)_\mu q^\mu}\ ;\quad\alpha,\beta\in\Bbb R^4\ .\eqno(3.20)$$
Such realizations will be termed {\it regular\/}.

We shall see in the next section that there is a well defined $C^*$--algebra ${\Cal E}$  whose non--degenerate
representations are in one--to--one correspondence with the regular realizations. ${\Cal E}$ describes our Quantum
Spacetime in the sense that it may be thought of as the space of continuous functions on the Quantum Spacetime
vanishing at infinity.

The situation is therefore analogous to quantum mechanics where we have a $C^*$--algebra  whose representations are
in one--to--one correspondence with regular realizations of the canonical commutation relations. Furthermore, just
as in quantum mechanics, there are other realizations and thus the possibility of defining $C^*$--algebras
describing other natural classes of representations and hence other Quantum Spacetimes. Taking quantum mechanics as
our guide, these possibilities will be ignored here.

An important role will be played by the {\it joint spectrum\/} of the commuting selfadjoint  operators
$Q_{\mu\nu}$. By (3.9) it is included in the set $\Sigma$ of all antisymmetric real 2--tensors $\sigma$ such that
$$\sigma_{\mu\nu}\sigma^{\mu\nu}=0\ ;\quad{1\over4}\,\sigma_{\mu\nu}(*\sigma)^{\mu\nu}=\pm1\ ,
\eqno(3.21)$$
i.e., writing $\sigma=(\vec e,\vec m)$; $e_j=\sigma_{0j}$, $m_j=\sigma_{lk}$, where $(j,l,k)$ is  a cyclic
permutation of $(1,2,3)$, such that
$$\vec e{\,}^2=\vec m^2\ ;\quad(\vec e\cdot\vec m)^2=1\ .\eqno(3.22)$$
We have $\Sigma=\Sigma_+\cup\Sigma_-$, where
$$\Sigma_\pm=\{\sigma=(\vec e,\vec m)/\vec e{\,}^2=\vec m^2\ ,\quad\vec e\cdot\vec m=\pm1\}\ .\eqno(3.23)$$

We can introduce the (Lorentz frame dependent) Euclidean norm on $\Sigma$ by
$$\|\sigma\|^2\equiv{1\over2}\,\sum_{\mu<\nu}\sigma^2_{\mu\nu}={1\over2}\,(\vec e{\,}^2+
\vec m^2)=\vec e{\,}^2=\vec m^2\ .$$
Then $\|\sigma\|\geq1$ for each $\sigma\in\Sigma$ and the unit sphere
$\Sigma^{(1)}$ of $\Sigma$ is  the compact
manifold
$$\Sigma^{(1)}=\{\sigma\in\Sigma/\|\sigma\|=1\}=\{\sigma=(\vec e,\vec m)\in\Sigma/\vec e=
\pm\vec m\}=\Sigma^{(1)}_+\cup\Sigma^{(1)}_-\ ,$$
i.e. $\Sigma^{(1)}_\pm$ is homeomorphic to $S^2$.

If the four selfadjoint operators $q_0,\dots,q_3$ obey the Quantum Conditions (3.9), (3.10) and  $\omega$ is a
state in the domain of the $Q_{\mu\nu}$'s then, by the Spectral Theorem, $\omega$ defines a regular probability
measure $\mu_\omega$ on $\Sigma$ s.t.
$$\omega(f(Q))=\int_\Sigma f(\sigma)d\mu_\omega(\sigma)\ ,\quad f\in{\Cal C}_0(\Sigma)\ .$$\medskip

\noindent{\bf 3.4 Proposition}. {\sl Under the hypothesis of {\rm Theorem 1}, we have
$$\Sigma_\mu(\Delta_\omega q_\mu)^2\geq\sqrt2\int_\Sigma(\|\sigma\|^2+1)^{1/2}d\mu_\omega(\sigma)\ .$$}\medskip

\noindent{\it Proof\/}. By (3.19) it sufficies to prove the above inequality for a state $\omega$ which is  pure on
the centre. $\mu_\omega$ is then a Dirac measure at a point $\sigma\in\Sigma$ and
$|\omega(Q_{\mu\nu})|=|\sigma_{\mu\nu}|$. Let $q'_\mu=\pi_\omega(q_\mu)-\omega(q_\mu)\cdot I$. For $\vec a\in\Bbb
R^3$ we set $\vec a\cdot\vec q\,'=\Sigma^3_{i=1}a_iq'_i$ and find the commutation relations
$$[q'_0,\vec a\cdot\vec q\,']=i\vec a\cdot\vec e\cdot I\ ,\quad[\vec a\cdot\vec q\,',\vec b\cdot\vec q\,']= i(\vec
a\times\vec b)\cdot\vec m\cdot I\ .$$ Now let $\{\vec a,\vec b,\vec c\}$ be an orthonormal basis of $\Bbb R^3$ with
$\vec a=\vec b\times\vec c$.

Then
$$\eqalign{
\sum_\mu(\Delta_\omega(q_\mu))^2&=\omega(q{'}^2_0+(\vec a\cdot\vec q\,')^2+(\vec b\cdot\vec q\,')^2+
(\vec c\cdot\vec q\,')^2)\cr
&\geq2\Delta_\omega(q'_0)\Delta_\omega(\vec a\cdot\vec q\,')+2\Delta_\omega(\vec b\cdot\vec q\,')
\Delta_\omega(\vec c\cdot\vec q\,')\cr
&\geq|\omega([q'_0,\vec a\cdot\vec q\,'])|+|\omega([\vec b\cdot\vec q\,',\vec c\cdot\vec q\,'])|\cr
&=|\vec a\cdot\vec e|+|(\vec b\times\vec c)\cdot\vec m|\geq|\vec a\cdot(\vec e\pm\vec m)|\ ,\cr}$$
for $\sigma\in\Sigma_+$ respectively.

The maximum over $\vec a$ is attained for $\vec a={\vec e\pm\vec m\over\|\vec e\pm\vec m\|}$, hence
$$\Sigma_\mu(\Delta_\omega(q_\mu))^2\geq\|\vec e\pm\vec m\|=\sqrt{2(\|\vec e\|^2+1)}=
\sqrt{2(\|\sigma\|^2+1)}\eqno\square$$

The last proposition sheds some light on the role of the manifold $\Sigma$. Very accurate measurements of  the
$q_\mu$'s select states $\omega$ for which $\mu_\omega$ is essentially concentrated on $\Sigma^{(1)}$. In generic
units, the unit sphere $\Sigma^{(1)}$ becomes the doubled sphere of radius $\lambda^2_P$.

As we will discuss in the next section, if we consider generic states, the manifold $\Sigma$ survives in the 
classical limit $\lambda_P\to0$. But if we limit ourselves to very well localized states, the effect of $\Sigma$
will be not directly visible in that limit. In the next section we will describe explicitly states with optimal
localization where the quantity $\Sigma_\mu(\Delta_\omega q_\mu)^2$ actually reaches its minimal possible value.

We collect here some easy results on the manifold $\Sigma$ which are either obvious or proved in Appendix B.\medskip

I. $\Sigma$ is a homogeneous space of the full Lorentz group for the action
$$\Lambda\in L\ ,\quad\sigma\in\Sigma\to\Lambda\sigma\Lambda^T=\sigma'\in\Sigma\ ,$$
$$\sigma'_{\mu\nu}=\Lambda_{\mu}^{\mu'}\Lambda_{\nu}^{\mu'}\sigma_{\mu'\nu'}\ .\eqno(3.24)$$

Under the action (3.24), $\Sigma_\pm$ are $L^\uparrow_+$--homogeneous spaces; $\Sigma_\pm$ are connected.  If $\vec
e=\vec m$, the stabilizer of $\sigma=(\vec e,\vec m)$ in $L^\uparrow_+$ consists precisely of boosts along $\vec e$
combined with rotations around $\vec e$. If $\vec e=\vec m$ is chosen as the third axis, the stabilizer of $\sigma$
is the image of the subgroup $D$ of diagonal matrices under the usual covering map $SL(2,\Bbb C)\to L^\uparrow_+$.
Hence we have homeomorphisms and isomorphisms of $L^\uparrow_+$--homogeneous spaces:
$$\Sigma_+\sim\Sigma_-\sim SL(2,\Bbb C)/D\ .\eqno(3.25)$$\smallskip

II. As a topological space, $\Sigma_+$ is homeomorphic to $TS^2$, the tangent bundle of the unit sphere  in $\Bbb
R^3$. The two--sphere $S^2$, naturally embedded in $TS^2$, corresponds precisely to $\Sigma^{(1)}_+$, the Euclidean
unit sphere of $\Sigma_+$. In particular, $S^2$ is a deformation retract of $\Sigma_+$.\smallskip

III. There are Borel sections for the map of $L^\uparrow_+$ onto $\Sigma_+:\Lambda\in L^\uparrow_+\to
\Lambda\sigma_0\Lambda^T$, $\sigma_0$ being a given point in $\Sigma_+$; we can choose such a section
$$\sigma\in\Sigma_+\to\Lambda_\sigma\in L^\uparrow_+\ ,$$
$$\Lambda_\sigma\sigma_0\Lambda^T_\sigma=\sigma\ ,\qquad\sigma\in\Sigma_+\ ,\eqno(3.26)$$
to be continuous on the complement of a closed set $N$ with zero quasi--invariant measure. We can  choose two such
sections $\sigma\to\Lambda_\sigma$, $\sigma\to\Lambda'_\sigma$, such that $N\cap N'=\emptyset$, each of which can
be written in the form
$$\Lambda_\sigma=L_\sigma R_\sigma\ ,\qquad\sigma\in\Sigma_+\ ,\eqno(3.27)$$
where $L_\sigma$ is a boost, $R_\sigma$ a rotation and $\sigma\in\Sigma_+\to L_\sigma\in L^\uparrow_+$  is
continuous.

Choosing $\sigma_0\in\Sigma_+$ s.t. $\sigma_0=(\vec e,\vec m)$, $\vec e=\vec m=(0,1,0)$, we have
$$(\sigma^{\mu\nu}_0)=\pmatrix
0&-I\cr
I&0\endpmatrix\equiv s\ ,\eqno(3.28)$$
where $I$ is the unit $2\times2$ matrix. Thus the symplectic form $\alpha$, $\beta\in\Bbb R^4\to\alpha_\mu
\sigma^{\mu\nu}_0\beta_\nu$ can be written as $-Im(\tilde\alpha,\tilde\beta)$, where
$\tilde\alpha=(\alpha_0+i\alpha_2,\alpha_1+i\alpha_3)\in\Bbb C^2$. In particular, $\sigma_0$ and hence any
$\sigma\in\Sigma$, induces a non--degenerate symplectic form and is an invertible matrix.

A regular realization with $Q^{\mu\nu}=\sigma^{\mu\nu}_0\cdot I$ 
is given by the Schr\"odinger operators  in the
plane, $Q_j=$ multiplication by $s_j$ and $P_j=-i{\partial\over\partial s_j}$, $j=1,2$, in $L^2(\Bbb R^2,d^2s)=H$,
by setting
$$q^{\sigma_0}_0=P_1,\ q^{\sigma_0}_1=P_2\ ;\quad q^{\sigma_0}_2=Q_1\ ;\quad q^{\sigma_0}_3=Q_2\ .
\eqno(3.29)$$

By von Neumann's uniqueness theorem, each regular irreducible realization is a (improper) {\it Lorentz  transform
$q^\sigma$ of $q^{\sigma_0}$, $\sigma=\Lambda\sigma_0\Lambda^T=\sigma_{0\Lambda}$\/}, for some $\Lambda\in L$. By
reduction theory every regular realization will be a direct integral of multiples of $q^\sigma$'s.

A {\it Poincar\'e--covariant\/} realization can be easily constructed as follows. Denoting by  $\overline
x\in\overline H$ and $\overline A$ the complex conjugates of an element $x$ of a Hilbert space $H$ and of a linear
operator $A$ on $H$, i.e. $\overline{Ax}=\overline{Ax}$, define
$$\tilde q^\sigma=q^\sigma\otimes I\hbox{ on }H\otimes\overline H\ ,$$
$$P_\mu=-((\sigma^{-1}q^\sigma)_\mu\otimes I+I\otimes(\sigma^{-1}\overline q^\sigma)_\mu)\ .\eqno(3.30)$$
It is easily checked that the $P_\mu$ are generators of a representation ${\Cal U}_\sigma$ of $\Bbb R^4$,  ${\Cal
U}_\sigma(a)=e^{iP_\mu a^\mu}$, which is unitary, strongly continuous and induces {\it translations\/} on $\tilde
q^\sigma$:
$${\Cal U}_\sigma(a)^{-1}\tilde q^\sigma_\mu{\Cal U}_\sigma(a)=\tilde q^\sigma_\mu+a_\mu\cdot I\ .
\eqno(3.31)$$
We can now define, on the Hilbert space ${\Cal H}=\int_\oplus H\otimes\overline H d\Lambda\cong H\otimes \overline
H\otimes L^2(L)$, the operators
$$q_\mu\equiv\int_\oplus(\Lambda q^\sigma)_\mu d\Lambda\ ;\eqno(3.32)$$
$${\Cal U}(a,I)\equiv\int_\oplus{\Cal U}_\sigma(\Lambda^{-1}a)d\Lambda\ ,\quad a\in\Bbb R^4\ ;
\eqno(3.33)$$
$${\Cal U}(0,\Lambda):({\Cal U}(0,\Lambda)x)(\Lambda')=x(\Lambda^{-1}\Lambda'),\Lambda,\Lambda'\in L,
x\in{\Cal H}\ ;\eqno(3.34)$$
$${\Cal U}(a,\Lambda)\equiv{\Cal U}(a,I){\Cal U}(0,\Lambda)\ .\eqno(3.35)$$
It is easily checked that (3.35) defines a unitary strongly continuous representation of the full  Poincar\'e group
and (3.32) a regular realization, where
$${\Cal U}(a,\Lambda)^{-1}q_\mu{\Cal U}(a,\Lambda)=(\Lambda q)_\mu+a_\mu\cdot I\ .\eqno(3.36)$$
By von Neumann uniqueness and uniqueness of the quasi--invariant regular measure class on $SL(2,\Bbb C)/D$,  every
regular realization, covariant in the sense of (3.36), will have to be quasi--equivalent to (3.32).

We close this section with a remark on our Quantum Conditions (3.9), (3.10). We could as well have required  that
the two invariant combinations of the central operators $Q_{\mu\nu}$ appearing in (3.9) coincide with generic real
multiples of the identity, say $Q_{\mu\nu}Q^{\mu\nu}=2b\cdot I$,
$\left({1\over4}\,Q_{\mu\nu}(*Q)^{\mu\nu}\right)^2=(a^2-b^2)\cdot I$, with $a\geq|b|$. The above discussion would
have shown that for each state $\omega$, the number $\eta\equiv\omega(\vec e{\,}^2)$, $\mu=\omega(\vec m^2)$ fulfil
$\mu\eta\geq a^2-b^2$, $\mu-\eta=2b$. Hence $\eta\geq a-b$, $\mu\geq a+b$, leading as above to the uncertainty
relations
$$\Delta q_0\Sigma\Delta q_j\geq{1\over2}\,(a-b)\ ,$$
$$\Sigma\Delta q_j\Delta q_k\geq{1\over2}\,(a+b)\ .$$
Thus our choice $b=0$, $a=1$ is the obvious symmetric choice in natural units.

\beginsection{4. Quantum Spacetime}

In this section we discuss the $C^*$--algebra describing Quantum Spacetime. This part of our  discussion is not
hard but still rather technical, and readers who are less mathematically minded might prefer to limit themselves to
the statements of the theorems and then proceed to the discussion of localization and the classical limit following
Theorem 4.2.

In view of the discussion towards the end of Sect. 3, we may expect that the $C^*$--algebra  describing the Quantum
Spacetime associated with the {\it regular realizations\/} of the Quantum Conditions (3.9), (3.10) (cf. Eq. (3.20))
is related to the norm--closed algebra generated by
$$\int f(\alpha)e^{i\alpha^\mu q_\mu}d^4\alpha\ ;\quad f\in L^1(\Bbb R^4,d^4\alpha)\ ,\eqno(4.1)$$
where $q_\mu$ is the realization (3.32).

In order to nail down the appropriate algebra, it is more instructive to follow von Neumann's approach  to
uniqueness and first introduce a Banach $*$--algebra ${\Cal E}_0$ associated with the regular realizations.

Let us define ${\Cal E}_0$ as the Banach space of continuous functions from $\Sigma$ to  $L^1(\Bbb R^4,d^4\alpha)$
vanishing at infinity, equipped with the product, *, and norm:
$$(f\times g)(\sigma,\alpha)=\int f(\sigma,\alpha')g(\sigma,\alpha-\alpha')e^{(i/2)\alpha_\mu
\sigma^{\mu\nu}\alpha'_\nu}d^4\alpha'\ ,\eqno(4.2)$$
$$(f^*)(\sigma,\alpha)=\overline{f(\sigma,-\alpha)}\ ,\eqno(4.3)$$
$$\|f\|_{0,1}=\sup_\sigma\|f(\sigma,\cdot)\|_1\eqno(4.4)$$
We define the action $\tau$ of ${\Cal P}$ on ${\Cal E}_0$ by
$$\tau_{(a,\Lambda)}(f)(\sigma,\alpha)=f(\Lambda^{-1}\sigma\Lambda^{-1T};\Lambda^{-1}\alpha)
e^{-ia_\mu\alpha^\mu}\det\Lambda\ ,\eqno(4.5)$$
where of course $\det\Lambda=\pm1$.

The commutative $C^*$--algebra ${\Cal C}(\Sigma)$ of bounded continuous functions on $\Sigma$ is  embedded in the
multiplier algebra of ${\Cal E}_0$ by
$$g\in{\Cal C}(\Sigma),f\in{\Cal E}_0\to gf\in{\Cal E}_0\ ,$$
$$(gf)(\sigma,\alpha)=g(\sigma)f(\sigma,\alpha)\ .\eqno(4.6)$$
Every non--degenerate representation $\pi$ of ${\Cal E}_0$ then determines a non--degenerate  representation
$\tilde\pi$ of ${\Cal C}(\Sigma)$ s.t. (cf. (4.6))
$$\tilde\pi(g)\pi(f)=\pi(gf)=\pi(f)\tilde\pi(g)\eqno(4.7)$$
and $\pi$ is irreducible only if $\exists\,\sigma\in\Sigma$ s.t.
$$\tilde\pi(g)=g(\sigma)\cdot I\ .\eqno(4.8)$$

By von Neumann uniqueness, the irreducible representations $\pi$ fulfilling (4.8) for a fixed  $\sigma\in\Sigma$
are all equivalent to one another and to the representation of ${\Cal E}_0$ determined by $q^\sigma$
$$f\in{\Cal E}_0\to\int d^4\alpha f(\sigma,\alpha)e^{i\alpha^\mu q^\sigma_\mu}\ ,\eqno(4.9)$$
where (cf. (3.29))
$$q^\sigma=\Lambda q^{\sigma_0}\quad\hbox{if}\quad\sigma=\Lambda\sigma_0\Lambda^T\ ,\quad\Lambda \in L\
.\eqno(4.10)$$\medskip

\noindent{\bf 4.1 Theorem}. {\sl There exists a unique $C^*$--norm on ${\Cal E}_0$. The completion  ${\Cal E}$ is
the $C^*$--algebra associated with a trivial continuous field of elementary algebras on $\Sigma$, i.e. it is
isomorphic to ${\Cal C}_0(\Sigma,{\Cal K})$, where ${\Cal K}$ is the $C^*$--algebra of all compact operators on a
fixed separable infinite dimensional Hilbert space.}\bigskip

\noindent{\it Proof\/}. The representations (4.9), (4.10), for $\sigma\in\Sigma$, form a separating family,  hence
${\Cal E}_0$ admits a $C^*$--norm. Let $\|\ \|$ denote the maximal $C^*$--norm and ${\Cal E}$ the associated
completion. A non--zero representation $\pi$ of ${\Cal E}$ is irreducible iff $\pi|{\Cal E}_0$ is irreducible, and
hence iff $\pi|{\Cal E}_0$ is unitarily equivalent to the representation given by (4.9), (4.10) for some
$\sigma\in\Sigma$.

Therefore, for each $\sigma\in\Sigma$ there is an irreducible representation $\pi_\sigma$ of ${\Cal E}$  satisfying
(4.8) which is unique up to equivalence and we have
$$\pi_\sigma({\Cal E})={\Cal K}({\Cal H}_{\pi_\sigma})\ ,\quad\sigma\in\Sigma\ .\eqno(4.11)$$
If we embed ${\Cal E}_0$ into the Banach $*$--algebra $\tilde{\Cal E}_0$ of bounded continuous functions  from
$\Sigma$ to $L^1(\Bbb R^4)$ with product, *, and norm given by (4.2), (4.3), (4.4), then clearly ${\Cal
E}_0\subset\tilde{\Cal E}_0\subset M({\Cal E}_0)$, the multiplier algebra of ${\Cal E}_0$. The maximal $C^*$--norm
$\|\cdot\|$ on ${\Cal E}_0$ induces a $C^*$--norm, still denoted by $\|\cdot\|$, on $\tilde{\Cal E}_0$. If
$\tilde{\Cal E}$ is the completion of $\tilde{\Cal E}_0$ in that norm, we have
$${\Cal E}\subset\tilde{\Cal E}\subset M({\Cal E})\eqno(4.12)$$
and, for each $\sigma\in\Sigma$, $\pi_\sigma({\Cal E})=\pi_\sigma(\tilde{\Cal E})={\Cal K}({\Cal H}_{\pi_\sigma})$.

We will show that the bundle $\{\pi_\sigma(\tilde{\Cal E});\sigma\in\Sigma\}$ equipped with the continuous  fields
$\{\{\pi_\sigma(A):\sigma\in\Sigma\}$, $A\in\tilde{\Cal E}\}$ is {\it a trivial continuous field\/} of elementary
algebras. ${\Cal E}$ is then the associated $C^*$--algebra, i.e. the $C^*$--algebra of continuous fields vanishing
at infinity, and is isomorphic to ${\Cal C}_0(\Sigma,{\Cal K})$.

To this end it suffices to show that our bundle is locally trivial and that there is a continuous field of 
one--dimensional projections (cf. [21, ch. 10, 7.6, 7.15, 8.4]).

We therefore choose the representations $\pi_\sigma$, $\sigma\in\Sigma$ in an appropriate way. For each 
$\sigma\in\Sigma$, let $f_0(\sigma,\cdot)\in L^1$ be the von Neumann element in the fibre of ${\Cal E}_0$ at
$\sigma$ corresponding to the selfadjoint projection on the ground state of the harmonic oscillator. Namely, if
$\sigma\in\Sigma\to\Lambda_\sigma\in L$ is a section as in Sect. 3, III (cf. eq. (3.29) and II, Appendix B), we
define $f_0$ by
$$f_0(\sigma,\alpha)={1\over(2\pi)^4}\,e^{-{1\over4}\,(\alpha,\alpha)_\sigma}\ ,\quad\sigma\in\Sigma\ ,
\quad\alpha\in\Bbb R^4\ ,\eqno(4.13)$$
$$(\alpha,\alpha)_\sigma=(\Lambda^{-1}_\sigma\alpha,\Lambda^{-1}_\sigma\alpha)=(L^{-1}_\sigma
\alpha,L^{-1}_\sigma\alpha)\ ,\quad\sigma\in\Sigma,\ \alpha\in\Bbb R^4\ ,\eqno(4.14)$$
where $(\alpha,\alpha)=\sum^3_{i=0}\alpha^2_i$ and we used the specific form of our section $\Lambda_\sigma=
L_\sigma R_\sigma$, where $R_\sigma$ is a rotation thus leaving the Euclidean scalar product invariant.

Since the Lorentz group acts continuously on $L^1(\Bbb R^4):f\in L^1(\Bbb R^4)\to f_L$: $f_L(\alpha)=f(L^{-1}
\alpha)$, and $\sigma\in\Sigma\to L_\sigma$ is continuous (cf. Appendix B), $\sigma\to f_0(\sigma,\cdot)\in L^1$ is
continuous, where $f_0(\sigma,\cdot)=f_0(\sigma_0,\cdot)_{L_\sigma}$.

Now $\pi_\sigma(f_0)$ is a selfadjoint projection of rank 1 [22], so that $\{\pi_\sigma(f_0);
\sigma\in\Sigma\}$ is a continuous field of rank one projections.

Furthermore the following relation holds for each $\sigma\in\Sigma$:
$$(f_0\times f\times f_0)(\sigma,\cdot)=\omega_\sigma(f)f_0(\sigma,\cdot),\ f\in\tilde{\Cal E}_0\ ,
\eqno(4.15)$$
and defines a state $\omega_\sigma$ on $\tilde{\Cal E}$. From now on let $(\pi_\sigma,{\Cal H}_\sigma, \xi_\sigma)$
be the GNS construction for $\omega_\sigma$. We have only to show that our continuous field is locally trivial.

Let $\Lambda_1$, $\Lambda_2$ be two sections as in III, Sect. 3, continuous on $\Sigma_1$, $\Sigma_2$ resp.  where
$\Sigma_1$, $\Sigma_2$ is an open covering of $\Sigma$ (cf. Appendix B). Define an isomorphism $\rho_i$ of ${\Cal
C}_0(\Sigma_i,L^1)$ regarded as the tensor product of ${\Cal C}_0(\Sigma_i)$ and the fibre of ${\Cal E}_0$ at
$\sigma_0$ (cf. (4.2)) onto ${\Cal C}_0(\Sigma_i,L^1)$ regarded as a subalgebra of ${\Cal E}$ by
$$\rho_i(f)(\sigma,\alpha)=f(\sigma,\Lambda^{-1}_{i\sigma}\alpha)\ ;\eqno(4.16)$$
one checks that $\rho_i$ is a $*$--isomorphism defining a trivialization of our field on $\Sigma_i$. 

\hfill $\square$\medskip

\noindent{\it Remark\/}. The triviality of our continuous field would follow directly if we could define a  map,
continuous in the appropriate sense, assigning to each $\sigma\in\Sigma$ a regular solution $\tilde q^\sigma$ s.t.
$[\tilde q^\sigma_\mu,\tilde q^\sigma_\nu]=i\sigma_{\mu\nu}\cdot I$. We might tentatively define $\tilde q^\sigma$
as linear combinations of the fixed choice $q^{\sigma_0}$ of Eq. (3.29) but this approach fails, because the bundle
$\Lambda\in L\to\sigma_\Lambda\in\Sigma$ is non--trivial and the symplectic group is not simply connected. However,
we have the further possibility of deforming $q^{\sigma_0}$ to $Uq^{\sigma_0}U^{-1}$, where $U$ is an element of
the unitary group of $H$ using the fact that this group is contractible (cf. [21, Chap. 10]).

It follows from the proof of Theorem 4.1 that each $C^*$--seminorm on ${\Cal E}$ has the form
$$\|f\|_S=\sup_{\sigma\in S}\|\pi_\sigma(f)\|\ ,$$
for some closed subset $S$ of $\Sigma$. Since $\Sigma$ is a homogeneous space, the unique $C^*$--norm  is also {\it
the only non--zero $\tau$--invariant $C^*$--seminorm\/}, and $\tau$ extends to a strongly continuous action, still
denoted by $\tau$, of the full Poincar\'e group by automorphisms of ${\Cal E}$.

The quasiequivalence classes of representations of ${\Cal E}$ are labelled by the regular measure  classes on
$\Sigma$. A covariant representation $\pi$ of ${\Cal E}$ obviously yields a covariant representation $\tilde\pi$ of
${\Cal C}_0(\Sigma)$ and is hence associated with a quasi--invariant regular measure on $\Sigma$. Since there is
only one invariant measure class, it follows that {\it there is only one quasiequivalence class of representations
$\pi$ of ${\Cal E}$ s.t. $(\pi,{\Cal U})$ is a covariant representation of $({\Cal E},\tau)$\/} for some
representation ${\Cal U}$ of the full Poincar\'e group.

The $C^*$--algebra ${\Cal E}$ of Theorem 4.1 does indeed describe the Quantum Spacetime associated  with the class
of regular solutions of the Quantum Conditions (3.9), (3.10), in the sense of the previous section; namely:\bigskip

\noindent{\bf 4.2 Theorem}. {\sl For each $f_1\in{\Cal C}_0(\Sigma)$ and $f_2\in L^1(\Bbb R^4)$ let  $f_1\otimes
f_2\in{\Cal E}_0\subset{\Cal E}$ be the function $\sigma\in\Sigma\to f_1(\sigma)f_2\in L^1(\Bbb R^4)$. We can
define selfadjoint operators $q_\mu$, $Q_{\mu\nu}$, affiliated to ${\Cal E}$ with support of definition I (Appendix
A), by setting, for each nondegenerate representation $\pi$ of ${\Cal E}$,
$$\pi(f_1\otimes f_2)=f_1(\pi(Q))\int f_2(\alpha)e^{i\alpha_\mu\pi(q^\mu)}d^4\alpha\ ,$$
$$f_1\in{\Cal C}_0(\Sigma),\ f_2\in L^1(\Bbb R^4)\ .\eqno(4.17)$$
These operators obey the Quantum Conditions since
$$[q_\mu,q_\nu]^-=iQ_{\mu\nu}\ ,$$
the $Q_{\mu\nu}$ commute and joint spectrum of $\{Q_{\mu\nu},\mu,\nu=0,\dots,3\}=\Sigma$ (cf. IV, Appendix A).

\noindent Moreover, for each element $(a,\Lambda)$ of the full Poincar\'e group, we have
$$\tau^{-1}_{(a,\Lambda)}(q_\mu)=(\Lambda q+a\cdot I)_\mu\ ,\eqno(4.18)$$
where the automorphisms of ${\Cal E}$ act on selfadjoint operators affiliated to ${\Cal E}$  as described in
Appendix A III.}\bigskip

We will refrain from spelling out the easy proof of this theorem and limit ourselves to the  remark that, by the
relations defining $q_\mu$ and $Q_{\mu\nu}$, if two representations $\pi_1$, $\pi_2$ of ${\Cal E}$ fulfill
$\pi_1(q_\mu)=\pi_2(q_\mu)$, then $\pi_1(Q_{\mu\nu})=\pi_2(Q_{\mu\nu})$, and $\pi_1$, $\pi_2$ agree on a total
subset of ${\Cal E}_0$, hence $\pi_1=\pi_2$ on ${\Cal E}$.

In view of Theorem 4.1 we can rewrite Eq. (4.17) as an identity in ${\Cal E}$ involving the  selfadjoint operators
$q_\mu$, $Q_{\mu\nu}$ affiliated to ${\Cal E}$:
$$f_1\times f_2=f_1(Q)\int f_2(\alpha)e^{i\alpha_\mu q^\mu}d^4\alpha\ .\eqno(4.19)$$

We next discuss briefly the physical meaning of the state space ${\Cal S}({\Cal E})$.  We interpret each state
$\omega\in{\Cal S}({\Cal E})$ as specifying the localization of events. Positivity of $\omega$ and the commutation
relations prevent absolute precision in the localization, but there are states $\omega$ having optimal localization
properties compatible with Proposition 3.4. The associated measure $\mu_\omega$ on $\Sigma$ will be concentrated on
the unit sphere $\Sigma^{(1)}$, i.e. on the base $S^2\times\{\pm1\}$ of $TS^2\times\{\pm1\}\sim\Sigma$. The
associated operators $q^\sigma_\mu$, $\sigma\in\Sigma^{(1)}$, will be obtained from the solution (3.29), associated
to $\sigma_0:\sigma^{\mu\nu}_0=\pmatrix 0&-I\cr I&0\endpmatrix$, by an (improper) rotation.

As is well known from elementary Quantum Mechanics, the minimum of the quantity (cf. (3.29))
$$\Sigma_\mu(\Delta q^\sigma_\mu)^2=(\Delta Q_1)^2+(\Delta P_1)^2+(\Delta Q_2)^2+(\Delta P_2)^2$$
is actually 2, and is attained on states which are translates of the ground state of the Hamiltonian  $H$ of the
harmonic oscillator in 2 dimensions,
$$H={1\over2}\,(P^2_1+Q^2_1+P^2_2+Q^2_2)={1\over2}\,\Sigma_\mu(q^\sigma_\mu)^2\ .$$
Therefore, we can parametrize the states on ${\Cal E}$ with optimal localization by a vector $x$ in  Minkowski
space and a measure $\mu$ on $\Sigma$ {\it carried\/} by $\Sigma^{(1)}$ and define the associated state $\omega$ by
its restriction to ${\Cal E}_0$:
$$\omega(f)=\int d^4\alpha d\mu(\sigma)f(\sigma,\alpha)e^{i\alpha_\mu x^\mu-{1\over2}\,\Sigma_\mu
\alpha^2_\mu}\ ,$$
$$f\in{\Cal C}_0(\Sigma,L^1(\Bbb R^4))={\Cal E}_0\ ,\eqno(4.20)$$
so that $\omega$ is specified by the values of its normal extension $\tilde\omega$ to the multiplier algbra through
$$\tilde\omega(f(Q)e^{i\alpha_\mu q^\mu})=e^{i\alpha_\mu x^\mu-{1\over2}\,\Sigma_\mu\alpha^2_\mu}
\int f(\sigma)d\mu(\sigma)\ ,\quad\alpha\in\Bbb R^4,\ f\in{\Cal C}(\Sigma)\ .\eqno(4.21)$$
We close this section with a remark on the {\it classical limit\/} $\lambda_P\to0$.

\noindent In generic units, the twisting factor in the product (4.2) of ${\Cal E}_0$ takes the form
$$e^{{i\over2}\,\lambda^2_P\alpha_\mu\sigma^{\mu\nu}\alpha'_\nu}\ ,\eqno(4.22)$$
so that, as $\lambda_P\to0$, our algebra ${\Cal E}$ deforms to
$${\Cal C}_0(\Bbb R^4)\otimes{\Cal C}_0(\Sigma)={\Cal C}_0(\Bbb R^4)\otimes{\Cal C}(\{\pm1\})
\otimes{\Cal C}_0(\Sigma_+)\ ,$$
that is our non--commutative space deforms to the (commutative) space
$$\Bbb R^4\times\{\pm1\}\times\Sigma_+\ .$$
The factor (4.22) corresponds to writing the Quantum Conditions on $q_\mu$, in generic units, in the form
$$[q_\mu,q_\nu]=i\lambda^2_P\tilde Q_{\mu\nu}\ ,$$
$$\tilde Q_{\mu\nu}(\sigma)=\sigma_{\mu\nu}\ ,\qquad\sigma\in\Sigma\ ,\eqno(4.23)$$
where the $\tilde Q_{\mu\nu}$ defined this way are selfadjoint operators affiliated to ${\Cal C}_0(\Sigma)$.

Equation (4.23) implies that the large dilation limit and the classical limit coincide. If we  were prepared to
{\it violate\/} this condition we could write
$${1\over4}\,[q_0,\dots,q_3]^2=\lambda^8_P\cdot I\ ,\qquad Q_{\mu\nu}Q^{\mu\nu}=0$$
in place of (4.23) and the $\lambda_P\to0$ limit would give a {\it dilation covariant Quantum  Spacetime\/},
defined by
$$[q_\mu,q_\nu]=i{\Cal R}_{\mu\nu}\ ,$$
$${\Cal R}_{\mu\nu}(\sigma)=\sigma_{\mu\nu}\ ,\quad\sigma\in\Sigma_0\ .$$
$\Sigma_0$ is the set of all real antisymmetric 2--tensors such that
$$\sigma_{\mu\nu}\sigma^{\mu\nu}=\sigma_{\mu\nu}(*\sigma)^{\mu\nu}=0\ .$$
Now $\Sigma_0$ is {\it connected\/} and is a single orbit under $L^\uparrow_+$ of any  $\sigma\in\Sigma_0$; for
each $\sigma\in\Sigma_0$, $\sigma=(\vec e,\vec m)$ with $\vec e{\,}^2=\vec m^2$, $\vec e\cdot\vec m=0$. For each
$\lambda>0$, an appropriate boost along $\vec e\times\vec m$ will change $\sigma=(\vec e,\vec m)$ to
$\lambda\sigma$. The symplectic form defined by $\sigma\in\Sigma_0$ is now degenerate, $\sigma\cong\pmatrix 0&-1\cr
1&0\endpmatrix\oplus\pmatrix 0&0\cr 0&0\endpmatrix$.

\beginsection{5. Calculus on the Quantum Spacetime}

Let $\tilde{\Cal E}$ denote, as before, the $C^*$--algebra of continuous bounded functions from  $\Sigma$ to ${\Cal
K}$, and $Z\subset M(\tilde{\Cal E})$ the Abelian $C^*$--algebra of bounded complex continuous functions on
$\Sigma$.

With $f\in L^1(\Bbb R^4)$ let $\hat f={\Cal F}f\equiv g$ and $f=\overline{\Cal F}g=\check g$  denote the Fourier
transform and its inverse.

For each $f\in{\Cal F}L^1(\Bbb R^4)$ we can define the function $f(q)$ of the quantum coordinates  $q_\mu$ as an
element of $\tilde{\Cal E}$ by
$$f(q)\equiv\int\check f(\alpha)e^{iq_\mu\alpha^\mu}d^4\alpha\ ,\quad f\in{\Cal F}L^1(\Bbb R^4)\ .\eqno(5.1)$$
Spacetime translations act as automorphisms $\tau_a$, $a\in\Bbb R^4$, such that (cf. Sect. 3,4)
$$\tau_a(f(q))=f(q-a\cdot I)\ ,\qquad a\in\Bbb R^4\ .\eqno(5.2)$$
We can now define spacetime derivatives as in Minkowski space as minus the infinitesimal generator of 
translations, i.e.
$$\partial_\mu f(q)\equiv{\partial\over\partial a_\mu}\,f(q+a\cdot I)|_{a=0}\ .\eqno(5.3)$$
If we take the product of $n$ operators $f_1(q),\dots,f_n(q)$ defined as in (5.1), we get another  function of $q$
which on the one side is distinct from the pointwise product $f_1\dots f_n$ evaluated on $q$ and on the other is no
longer $\Bbb C$--valued but $Z$--valued in general. Both facts are apparent from the explicit formula which is an
immediate consequence of (5.1):
$$f_1(q)\dots f_n(q)=(\check f_1\times\dots\times\check f_n)^\wedge(q)\ ,\eqno(5.4)$$
where the twisted convolution $\times$ depends on $Q_{\mu\nu}$:
$$(h\times h')(\alpha)=\int h(\alpha')h'(\alpha-\alpha')e^{i/2\alpha Q\alpha'}d^4\alpha\ ,$$
$$\alpha Q\alpha'\equiv\alpha_\mu Q^{\mu\nu}\alpha'_\nu\ ;\quad h,h'\in L^1(\Bbb R^4)\ ,\eqno(5.5)$$
so that
$$(\check f_1\times\dots\times\check f_n)(k)=\int d^4k_1\dots d^4k_n\delta^{(4)}(k-\Sigma k_j)
\cdot e^{i/2\Sigma_{j<l}k_jQk_l}$$
$$\times\check f(k_1)\dots\check f(k_n)\ .\eqno(5.6)$$
In order to develop Quantum Field Theory on the Quantum Spacetime and to apply the conventional  perturbation
methods, it will be important to define the quantum analogues of the positive linear functional on $L^1(\Bbb
R^4)\cap{\Cal F}L^1(\Bbb R^4)$ given by the total integral and the integral over space at a fixed time $t$:
$$f\in L^1\cap{\Cal F}L^1\to\int f(x)d^4x=\check f(0)\ ,\eqno(5.7)$$
$$f\in L^1\cap{\Cal F}L^1\to\int_{x_0=t}f(t,\vec x)d^3x=\int e^{itk_0}\check f(k_0,\vec0)dk_0\ .
\eqno(5.8)$$
Define the $Z$--valued trace $Tr$ on $\tilde{\Cal E}$ by
$$(Tr(X))(\sigma)=tr X(\sigma)\ ,\quad\sigma\in\Sigma\ ,\eqno(5.9)$$
where $tr$ is the usual trace on $L^1({\Cal B}({\Cal H}))\subset{\Cal K}({\Cal H})$.

If $f\in L^1(\Bbb R^4,Z)\cap{\Cal F}L^1(\Bbb R^4,Z)$ the formal analogue of (5.7), (5.8) can be defined by
$$\int d^4qf(q)\equiv\int f(x)d^4x=\check f(0)=Tr f(q)\ ,\eqno(5.10)$$
$$\eqalignno{
\int_{q_0=t}f(q)d^3q&\equiv\int e^{ik_0t}\check f(k_0,\vec0)dk_0\cr
&=\lim_mTr(f_m(q)^*f(q)f_m(q))\ .&(5.11)\cr}$$
Note that (5.10), (5.11) are consistent since
$$\int dt\int_{q_0=t}f(q)d^3q=\int f(q)d^4q\ .$$
Positivity can be established either directly, using the first line of (5.11) as a definition and  (5.5) to show that
$$\int_{q_0=t}f(q)^*f(q)d^3q\geq0\ ,$$
where of course $f(q)^*=\overline f(q)$ as a consequence of the definition (5.1); or else by showing  that the
second equality in (5.11) does hold for an appropriate sequence $f_m$. To this end it suffices to choose real
functions $f_m$ such that $f^2_m$ approximates the constant function one on space times the Dirac measure at time
$t$. Then, using (5.6) for $n=3$, we get
$$\eqalignno{
Tr f_m(q)^*f(q)f_m(q)&=Tr f_m(q)\overline f_m(q)f(q)=(\check f_m\times\check f_m\times f)(0)\cr
&=\int d^4k_1d^4k_2d^4k_3\delta^{(4)}(\Sigma k_j)\cdot e^{(i/2)\Sigma_{j<l}k_jQk_l}\check f_m(k_1)
\check f_m(k_2)\check f(k_3)\cr
&=\int d^4k_1d^4k_2e^{(i/2)k_1Qk_2}\check f_m(k_1)\check f_m(k_2)\check f(-(k_1+k_2))\ ,&(5.12)\cr}$$
where $\check f_m*\check f_m$ approximates $\delta^{(3)}(\vec k)\cdot e^{-ik_0t}$, so that $e^{(i/2)k_1Qk_2}$  can
be replaced by $I$ in the limit, since $Q_{00}=0$, and the expression (5.12) approximates $\int dk_0\check
f(k_0,\vec0)$ $e^{-ik_0t}$, as desired. The positivity of the weight (5.11) can be interpreted in the light of our
spacetime uncertainties which must hold for any {\it positive\/} functional and are compatible with absolute
precision in the measurement of time together with complete lack of knowledge of the space coordinates. On the
contrary, the functional $f(q)\to f(x)$, $x\in\Bbb R^4$, for instance, is {\it not\/} positive on $\tilde{\Cal E}$.

In order to obtain a $\Bbb C$--valued functional we must integrate over $Z$ too. If the result is to agree  with
the classical definition for functions taking values in $\Bbb C\cdot I\subset Z$, we must use a normalized state on
$Z$. Unfortunately there is no obvious Lorentz invariant choice since $L^\uparrow_+$ is not amenable. Motivated by
our discussion of localization, we will choose the state defined by integration over the unit sphere $\Sigma^{(1)}$
with the normalized Lebesgue measure $d\sigma$. Of course the result will be rotation but not Lorentz invariant.

We first analyze the $Z$--valued weight $\int_{q_0=t}d^3q$ in more detail. In fact, in order be able  to define an
interaction Hamiltonian we will need expressions like
$$\int_{q_0=t}f_1(q)\dots f_n(q)d^3q\ .\eqno(5.13)$$
Combining (5.6) with (5.11) we see that (5.13) is given by
$$\int{\Cal D}_n(x_1,\dots,x_n;t)f_1(x_1)\dots f_n(x_n)d^4x_1\dots d^4x_n\ ,\eqno(5.14)$$
where the $Z$--valued kernel ${\Cal D}_n$ is given by
$${\Cal D}_n(x_1,\dots,x_n;t)=
\int dk_0d^4k_1\dots d^4k_ne^{-i\Sigma_jk_jx_j}e^{ik_0t}
\delta^{(1)}(k_0- \Sigma
k_{i0})\cdot $$ $$\delta^{(3)}
(\Sigma\vec k_i)e^{(i/2)\Sigma_{j<l}k_jQ_{k_l}}\ .\eqno(5.15)$$

The explicit expressions are (cf. Appendix C):
$${\Cal D}_{2n+1}(x_1,\dots,x_{2n+1};t)=\pi^{-4n}e^{2i\Sigma_{j\leq l\leq n}(x_{2j}-x_{2j-l})
Q^{-1}(x_{2l+1}-x_{2l})}\cdot$$
$$\delta\left(\sum^{2n+1}_{j=1}(-1)^j(x^0_j-t)\right)\ ,$$
and
$${\Cal D}_{2n+2}(x_1,\dots,x_{2n+2};t)=\pi^{-4n}{1\over2\pi}\,\int d\eta e^{i\eta\Sigma^{2n+1}_{j=1}
(-1)^j(x^0_j-t)}\cdot$$
$$e^{2i\Sigma_{j\leq l\leq n}(x_{2j}-x_{2j-1})Q^{-1}(x_{2l+1}-x_{2l})}\delta^{(4)}
\left(\sum^{n+1}_{j=1}(x_{2j}-x_{2j-1})-{1\over2}\,Q(\eta,\vec0)\right)\ .$$

The kernel ${\Cal D}_n$ describes the non--locality introduced by the quantum nature of spacetime as  a deviation
of (5.13) from the classical expression $\int f_1(\vec x, t)\dots f_n(\vec x,t)d^3x$. We limit ourselves to the
explicit expressions for $n=2,3$, which will be of particular importance in the next section. For $n=2$ we get
$$\int_{q_0=t}f_1(q)f_2(q)d^3q=(2\pi)^{-1}\int f_1(x)f_2(x+Q(\eta,\vec0))e^{-i\eta(t-x_0)}d^4xd
\eta\ .\eqno(5.16)$$
Note that integrating over $t$ gives
$$\int dt\int_{q_0=t}f_1(q)f_2(q)d^3q\equiv \int f_1(q)f_2(q)d^3q=\int f_1(x)f_2(x)d^4x\ ,$$
so that there is no nonlocal effect in the spacetime integral when $n=2$.

The nonlocal effects also disappear from the space integral at fixed  time if we restrict $f_1$, $f_2$  to be
solutions of the Klein--Gordon equation $(\square+m^2)f=0$ and evaluate the usual scalar product
$$\eqalign{
(f,g)&\equiv\int_{x_0=t}\overline f(x)\buildrel{\leftrightarrow}\over{\partial_0}g(x)d^3x\cr
&=\int_{x_0=t}\left({\overline{\partial f}\over\partial x_0}\,(x)g(x)-\overline{f(x)}
{\partial g\over\partial x_0}\,(x)\right)d^3x\ .\cr}$$
Indeed we have for any such $f$ and $g$ and for each $t$,
$$\int_{q_0=t}\overline{f(q)}\buildrel{\leftrightarrow}\over{\partial_0}g(q)d^3q=(f,g)\cdot I\ .\eqno(5.17)$$
To prove (5.17) it suffices to note that, by (5.16), the l.h.s. can be written as
$$(2\pi)^{-1/2}\int dx_0d\eta e^{-i\eta(t-x_0)}\int d^3xf(x)\buildrel{\leftrightarrow}\over {\partial_0}g_\eta(x)\
,\eqno(5.18)$$ where $g_n(x)\equiv g(x+Q(\eta,\vec0))$ is again a solution of the Klein--Gordon equation. Therefore
the $\int d^3x$ does not depend on $x_0$ and integrating over $x_0$ gives the Dirac measure in the variable $\eta$.
Integrating over $\eta$ replaces $g_\eta$ by $g_0=g$ and (5.17) equals $(f,g)\cdot I$ as desired.

The calculation for $n=3$ gives
$${\Cal D}_3(x_1,x_2,x_3;t)=\left({1\over\pi}\right)^4\delta(t-x_{30}-(x_{10}-x_{20}))
e^{2i(x_2-x_1)Q^{-1}(x_3-x_2)}\ ,\eqno(5.19)$$
so that
$$\int_{q_0=t}f_1(q)f_2(q)f_3(q)d^3q=\left({1\over\pi}\right)^4\int d^4ad^4b\int_{x_0=t+(b-a)_0}
f_1(x+a)f_2(x+b)f_3(x)$$
$$\times e^{2iaQ^{-1}b}d^3x\ .\eqno(5.20)$$
Thus the non--local effects are now visible in the full spacetime integral, too, as we see by  integrating (5.20)
over $t$:
$$\int f_1(q)f_2(q)f_3(q)d^4q=$$
$$\left({1\over\pi}\right)^4\int d^4ad^4bd^4xf_1(x+a)f_2(x+b)f_3(x)e^{2iaQ^{-1}b}\ .\eqno(5.21)$$
It is instructive to write (5.20) in {\it generic units\/}. To this end we must replace $a_\mu$ by 
$\lambda_Pa_\mu$, $b_\mu$ by $\lambda_Pb_\mu$ and $Q^{-1}$ by $\lambda^{-2}_PK$, where $a_\mu$, $b_\mu$ and
$K:\sigma\in\Sigma\to K(\sigma)=((\sigma^{\mu\nu})_{\mu,\nu=0,\dots,3})^{-1}$ are dimensionless.

We get
$$\int_{q_0=t}f_1(q)f_2(q)f_3(q)d^3q=$$
$${1\over\pi^4}\,\int d^4ad^4b\int_{x_0=t+\lambda_P(b-a)_0}f_1(x+\lambda_Pa)f_2(x+\lambda_Pb)
f_3(x)\cdot e^{2i(a,Kb)}d^3x\ .\eqno(5.22)$$
Integrating over the $t$ variable and setting
$$\beta=2\lambda^{-1}_PKb\ ;\qquad b=\lambda^{-1}_P{1\over2}\,Q\beta\ ,$$
we get
$$\int f_1(q)f_2(q)f_3(q)d^4q={1\over(2\pi)^2}\,\int d^4\beta d^4x\hat f_1(\beta)f_2
\left(x+{1\over2}\,\lambda^{-1}_PQ\beta\right)f_3(x)e^{-i\beta x}\ .$$
Note that, by the Lebesgue Dominated Convergence Theorem (recall that $f_i\in L^1\cap{\Cal F}L^1)$,  the limit of
the last expression as $\lambda_P\to0$ is
$${1\over(2\pi)^2}\,\int d^4\beta d^4x\hat f_1(\beta)f_2(x)f_3(x)e^{i\beta x}=\int d^4xf_1(x)
f_2(x)f_3(x)$$
as expected. The correct $\lambda_P\to0$ limit is also evident in (5.22) or more generally in (5.15).

Note that in (5.22) the non--local corrections to the classical formula are {\it at least quadratic in\/} 
$\lambda_P$: the linear term vanishes since
$$\eqalign{
\int a_\mu e^{2i(a,Kb)}d^4ad^4b&=-\int b_\mu e^{2i(a,Kb)}d^4ad^4b\cr
&={(2\pi)^4\over2^4}\,\int a_\mu\delta^{(4)}(a)d^4a=0\ .\cr}$$
This is a general feature: for each $n$, $\lambda_P$ appears in ${\Cal D}_n$ only through $Q$ (cf. (5.15))  and, in
generic units,
$$Q_{\mu\nu}(\sigma)=\lambda^2_P\sigma_{\mu\nu}\ .$$

Therefore the lowest order corrections to QFT on the usual Minkowski space will be at least {\it quadratic\/}  in
$\lambda_P$. Linear terms in $\lambda_P$ might appear only if we introduce {\it gravitational interactions\/}
explicitly in the theory.

Finally we discuss the $\Bbb C$--valued weight analogous to the classical space integral at fixed time, defined by
$$\int_{\Sigma^{(l)}}d\sigma\int_{q_0=t}f_1(q)\dots f_n(q)d^3q\ .\eqno(5.23)$$
This expression is easily obtained from (5.14), (5.13) evaluating the state $\int_{\Sigma^{(l)}}d\sigma$ over  $Z$.
A similar comment applies to the explicit formula (5.22) for $n=3$. Note that, for $\sigma\in\Sigma^{(1)}$, we have
$K(\sigma)=-\sigma$. For this holds when $\sigma=\sigma_0$, $(\sigma^{\mu\nu}_0)=s=\pmatrix 0&-I\cr
I&0\endpmatrix$, and $\Sigma^{(1)}$ is a single orbit under the improper rotation group. A standard computation
gives
$$\eqalignno{
\delta(a,b)&\equiv\int_{\Sigma^{(1)}}d\sigma e^{-2ia_\mu\sigma^{\mu\nu}b_\nu}\cr
&=-{1\over2}\,\left({\sin\gamma_+(a,b)\over\gamma_+(a,b)}\,+{\sin\gamma_-(a,b)\over\gamma_-(a,b)}\right)\ ,
&(5.24)\cr}$$
$$\gamma_\pm(a,b)=2\|a_0\vec b-b_0\vec a\pm\vec a\times\vec b\|\ .\eqno(5.25)$$
Thus replacing the exponential in (5.22) by $\delta(a,b)$ we get the desired explicit expression for (5.23)  in the
case $n=3$.

\beginsection{6. Towards Quantum Field Theory on the Quantum Spacetime}

In this section we will lay down our basic philosophy and take the first steps towards QFT on QST.  A more thorough
analysis will be deferred to a subsequent paper.

As required from the outset, the Poincar\'e group acts as a symmetry group on our QST (cf. Sect. 3 and 4).  This
fact allows us to retain Wigner's notion of elementary particles as being described by irreducible representations
of the covering group of the Poincar\'e group.

In accordance with one of our principles, the large scale limit of our QST agrees with its classical limit  and
yields the classical Minkowski space times an unobserved ghost manifold. This asymptotic behaviour is expected to
allow one to describe causality for QFT over QST as an asymptotic property  corresponding to the locality principle
[17] in the large scale limit.

The usual construction of asymptotic scattering states, which also involves a large scale limit, should  likewise
carry over.

When attempting a perturbative study of the $S$--matrix for QFT over QST, the first steps are to define  free
fields and interaction Hamiltonians.

Starting from Wigner's definition of particles, the usual Fock space construction yields a free field  associated
with an irreducible representation of the covering group of the Poincar\'e group.

There is no difficulty in evaluating this field on the QST, at least formally, as a function with values  in the
algebra spanned by the creation and destruction operators. For simplicity we consider a neutral scalar free field
$\phi(x)$. Evaluating on $q_\mu$ according to the rule (5.1) gives
$$\phi(q)={1\over(2\pi)^{3/2}}\,\int(e^{iq_\mu k^\mu}\otimes a(\vec k)+e^{-iq_\mu k^\mu}
\otimes a(\vec k)^*)d\Omega^+_m(\vec k)\ ,\eqno(6.1)$$
where $d\Omega^+_m(\vec k)={d^3\vec k\over2\sqrt{\vec k^2+m^2}}$ is the usual invariant measure over the  positive
energy hyperboloid of mass $m$:
$$\Omega^+_m=\{k\in\Bbb R^4/k_\mu k^\mu=m^2\ ,\qquad k_0>0\}\ .$$
In order to give a precise mathematical meaning to the formal expression (6.1) we may think of  a quantum field
over QST acting on a Hilbert space ${\Cal H}$ as a linear map, continuous in the appropriate topology, assigning to
test functions $f$ linear operators affiliated to the $C^*$--tensor product ${\Cal E}\otimes{\Cal B}({\Cal H})$ and
formally denoted by
$$f\to\int\phi(q+aI)f(a)d^4a\ .$$

It would be possible, and perhaps even natural, to define a different notion of free field over QST,  by letting
$\phi$ depend of $q$ {\it and\/} $\sigma\in\Sigma$.

This would amount to choosing the creation and annihilation operators $a$ in the CCR algebra over  $L^2(\Bbb
R^3\times\Sigma)$, where $\Sigma$ is equipped with the Lorentz invariant measure $d\sigma$ induced by a Haar
measure on $L$. Thus
$$\int_\Sigma f(\sigma)d\sigma\equiv\int_Lf(\Lambda\sigma_0\Lambda^T)d\Lambda\ ,$$
where $\sigma_0$ is a reference point in $\Sigma$. This approach would be closer in spirit to quantizing  wave
functions over QST and will be pursued elsewhere.

As a first simpler choice we will ignore here the possible $\sigma$--dependence of free fields.

From the expression (6.1) for our free field over QST and from the definition (5.3) of derivatives,  it is clear
that the Klein--Gordon equation holds
$$(\square+m^2)\phi(q)=0\ .\eqno(6.2)$$
Let $L\to{\Cal U}(L)$ be the unitary representation of the Poincar\'e group over ${\Cal H}$ defined  by the Fock
construction and $\alpha_L\equiv Ad\,{\Cal U}(L)$ the induced action on linear operators acting on ${\Cal H}$. The
relativistic covariance of our free field takes the form
$$\tau_L\otimes\alpha_L(\phi(q))=\phi(q)\ ,\eqno(6.3)$$
as a consequence of (4.19) and (6.1). In order to define ``local obsevables'' in this model, we will  consider
states $\omega$ over the $C^*$--algebra ${\Cal E}$ of QST as the analogues of test functions defining localization
data in spacetime. Pure states (with good localization properties) should play a role analogous to points in
classical space.

The free field defines a map from states $\omega\in{\Cal S}({\Cal E})$ to operators on ${\Cal H}$ by
$$\phi(\omega)\equiv\langle\omega\otimes id,\phi(q)\rangle\ ,\quad\omega\in{\Cal S}({\Cal E})\ .\eqno(6.4)$$
Introducing the test function $\psi_\omega$ associated to $\omega\in{\Cal S}({\Cal E})$ by
$$\psi_\omega(x)=\int e^{-ik_\mu x^\mu}\omega(e^{ik_\mu q^\mu}){d^4k\over(2\pi)^4}\ ,\eqno(6.5)$$
we can write the field operator $\phi$ evaluated at $\omega\in{\Cal S}({\Cal A})$ as the {\it usual free  field
smeared out with\/} $\psi_\omega$:
$$\phi(\omega)=\langle\phi,\psi_\omega\rangle=\int\phi(x)\psi_\omega(x)d^4x\ .\eqno(6.6)$$
Relations (6.5), (6.6) allow us to explore the locality properties of the free field over QST.  The commutator of
two ``values'' (6.4) of the field takes the form
$$[\phi(\omega),\phi(\omega')]=i\int\Delta(x-y)\psi_\omega(x)\psi_{\omega'}(y)d^4xd^4y\ .\eqno(6.7)$$
If we specialize $\omega$, $\omega'$ to translates of a given pure state with optimal localization of  the form
(4.20), say $\omega=\omega_a$, $\omega'=\omega_b$, we can compute the commutator (6.7) and study its asymptotic
properties. In this case, one easily sees that $\psi_\omega$, takes the form
$$\psi_{\omega_c}(x)=(2\pi)^{-2}e^{-{1\over2}\,\Sigma_\mu(x-c)^2_\mu}\ .\eqno(6.8)$$
The computation for the case $m=0$ can be carried out explicitly and exhibits those aspects of  causality which are
essential in the massive case, too.

We get, in generic units,
$$[\phi(\omega_a),\phi(\omega_b)]=\hfill$$
$$-i{1\over4\pi\|\vec a-\vec b\|}\,\cdot(8\pi\lambda^2_P)^{-1/2}\left(e^{-{1\over8\lambda^2_P}\,
(\|\vec a-\vec b\|-(a-b)_0)^2}-e^{-{1\over8\lambda^2_P}\,(\|\vec a-\vec b\|+(a-b)_0)^2}\right)I
\ .\eqno(6.9)$$
There are two important features of the expression (6.9). In spacelike directions it falls off  like a Gaussian,
hence faster than an exponential. If $a$, $b$ are kept fixed and we look at the limit $\lambda_P\to0$, the
expression (6.9) converges in the sense of distributions to the usual commutator function of the massless free
field, as expected.

We next discuss interaction Hamiltonians, briefly. It is important to note before--hand that the  {\it free\/}
Hamiltonian of a scalar free field of mass $m$ over QST can be written as the space integral of a ``quantum
density'' using the calculus of Sect. 5,
$$H=\int_{q_0=t}{\Cal H}(q)d^3q\ .\eqno(6.10)$$
To this end, note that the Hamiltonian of the ordinary neutral scalar free field can be written as
$$\eqalign{
H&=\int_{x_0=t}d^3x:{1\over2}\,\left({\partial\phi\over\partial x_0}\,(x)^2+(\vec\nabla\phi(x))^2+m^2
\phi(x)^2\right)\ :\cr &={1\over2}\,\int_{x_0=t}d^3x:\left({\partial\phi\over\partial x_0}\,(x)^2-{\partial^2\phi
\over\partial x^2_0}\,(x)\phi(x)\right):\ ,\cr}$$
where the double dots indicate Wick ordering. Here the Klein Gordon equation and integration by parts  have been
used. The relation (5.17), applied to the last integral, shows that (6.10) holds.

Consider now a more than bilinear interaction Hamiltonian density
$$H_I(x)=\lambda:\psi_1(x)\dots\psi_n(x):\ ,\eqno(6.11)$$
where $\psi_1,\dots,\psi_n$ are free field multiplets and summation over spin and internal degrees of  freedom is
implicit, $\lambda$ being a matrix of coupling constants.

It is natural to define an interaction Hamiltonian for QFT over QST associated with the interction (6.11)  by the
integral defined in Sect. 5 of the expression (6.11) evaluated
over the variables $q_\mu$:
$$H_I(t)\equiv\int_{\Sigma^{(1)}}d\sigma\int_{q_0=t}d^3q\lambda:\psi_1(q)\dots\psi_n(q):\ .\eqno(6.12)$$
By the results in Sect. 5 the expression (6.12) can be calculated and agrees with an {\it effective  nonlocal
Hamiltonian\/} defined by
$$H_I(t)=\int G(x_1,\dots,x_n;t)\lambda:\psi_1(x_1)\dots\psi_n(x_n):d^4x_1\dots d^4x_n\ ,$$
$$G(x_1,\dots,x_n;t)=\int _{\Sigma^{(1)}}d\sigma{\Cal D}_n(x_1,\dots,x_n;t)\ ,\eqno(6.13)$$
where the kernel ${\Cal D}_n$ is given by (5.15). In the important case of a trilinear interaction,  we get (cf.
(5.22))
$$H_I(t)={1\over\pi^4}\,\int d^4ad^4b\int_{x_0=t+\lambda_P(b-a)_0}\lambda:\psi_1(x+\lambda_Pa)
\psi_2(x+\lambda_Pb)\psi_3(x):\delta(a,b)d^3x\ ,\eqno(6.14)$$
where the kernel $\delta$ has been calculated in (5.24) and we have used generic units. Leaving out  the
integration over $d^3x$ in Eq. (6.14) we get, of course, the {\it effective nonlocal Hamiltonian density\/}.

Interacting fields can be tentatively defined in the following way: since $\phi$ is a solution of the 
Klein--Gordon equation, the form
$$\langle\phi,f\rangle_t=\int_\Sigma d\sigma\int_{q_0=t}d^3q(\phi(q)\partial_0f(q,\sigma)-
\partial_0\phi(q)f(q,\sigma))$$
is independent of $t$ for any solution $f\in M({\Cal E})$ of the Klein--Gordon equation. If $U(t,s)$  is the
unitary evolution operator with
$$U(t,s)U(s,r)=U(t,r)\ ,$$
$$U(t,t)=1\ ,$$
$${d\over dt}\,U(t,s)=iH_I(t)U(t,s)\ ,$$
we define the interacting field $\phi_{\text{int}}$ by
$$\langle\phi_{\text{int}},f\rangle_t=U(t,0)\langle\phi,f\rangle_0U(0,t)\ .$$
Since $\phi_{\text{int}}$ is not a solution of the wave equation, in general, the lefthand side depends on $t$.

By choosing suitable solutions of the wave equation for every $t$ and integrating over $t$ we determine 
$\phi_{\text{int}}$ on the whole state space of ${\Cal E}$.

Finally, we may follow the usual procedure and postulate LSZ asymptotic conditions which then lead to  the ordinary
perturbation expansion of the $S$--matrix.

As usual the $n^{\text{th}}$ order contribution can be calculated from the $n^{\text{th}}$ order  contribution to
the time--ordered function obtained substituting $H_I(t)$ in the formula
$${i^n\over n!}\,\int dt_1\dots dt_n(\Omega,T(A(x_1)\dots A(x_m)H_I(t_1)\dots H_I(t_n))\Omega)\ ,
\eqno(6.15)$$
where the ordering now refers to the $x_{1_0},\dots,x_{m_0}$, $t_1,\dots,t_n$ variables rather than  to the
integration variables in (6.13). This modifies the usual Feynmann rules. Due to the non--local character of our
effective interaction, for renormalizable theories, the renormalized perturbative expansion should agree with the
classical one for $\lambda_P=0$ and contain corrections of order $\lambda^2_P$ due to the quantum nature of
spacetime. We intend to give a more thorough discussion of these points elsewhere.

\beginsection{7. Outlook}

The choice of our QST was motivated by our principle that spacetime should have an operational meaning;  we
observed that this meaning would be destroyed by the gravitational collapse caused by preparing a very sharply
localized state.

We wish to point out that our uncertainty relations (2.8) and (2.9) appear as necessary but not a priori 
sufficient conditions to prevent gravitational collapse in such a process. Hence the Quantum Structure of our
spacetime might reflect only part of the necessary restrictions.

The very first steps to QFT  over QST outlined in Sect. 6 show that the quantum structure of our model  of
spacetime does lead to a smoothing of ultraviolet divergences. A more thorough analysis of the surviving
divergences and their renormalization in specific models is required (see [23] for related results).

Properties of the resulting theory which can be tested at a perturbative level (cluster properties,  convergence of
asymptotic states) can be a guide to explore properties of QFT over QST.

We wish to point out several other open problems.

Causality in QST has been mentioned here as an asymptotic property in the limit of vanishing Planck  length.
However, there might be an exact form, related to the causality properties of the free field over QST, expressing
the propagation of wave functions over QST. But even if such a tighter condition existed, it is not obvious whether
we can expect any stability under interactions.

We note that the $\Sigma$--dependence of interactions has been removed by a procedure which is not  Lorentz
invariant. Other approaches are possible, of course, e.g. one could treat the variables $\sigma\in\Sigma$ as random
variables. Or one could adopt a dynamical treatment introducing $\sigma$--dependent fields. We believe that a
satisfactory understanding of the role of $\Sigma$ should be possible if one were to incorporate gravitation in the
theory. Our QST ought to offer a more suitable basis for the formulation of Quantum Gravity, as it embodies part of
the limitations on the structure of spacetime determined by gravity, at least at a semiclassical level.

More generally, we could use the gauge principle as a natural way of introducing interactions between  fields over
the QST and Connes' non--commutative geometry [8] should provide the right framework here. The quantum nature of
spacetime has the effect of making QED into a non--commutative gauge theory, so that even the theory of
electro--magnetic fields without matter should become an interacting theory.

The large scale structure of the univese and the microstructure of spacetime relate to quite distinct  asymptotic
regimes. Yet it would be interesting and instructive to generalize our quantization procedure to a general curved
spacetime. The methods of deformation and geometric quantization could well apply here [24--26]. In this context,
it is quite clear that our formalism is not at all intended to exclude black hole formation at scales which are
larger than the Planck length.

Finally, it would be important to investigate whether a version of the Euclidean approach to QFT  is possible for
QFT over QST.

We hope to come back to these and related questions in a continuation of the present paper.

\beginsection{Appendix A}

Let {\got A} be a $C^*$--algebra with unit I. {\it A selfajdoint operator $A$ affiliated to\/}  {\got A} is defined
by a $*$--homomorphism (with unbounded support if $A$ is unbounded) of ${\Cal C}_0(\Bbb R)$ into {\got A}, denoted
$$f\in{\Cal C}_0(\Bbb R)\to f(A)\in\text{\got A}\ ,$$
whose support projection $E\in\text{\got A}^{**}$ is central. Equivalently, if $\{f_n;n=0,1,2, \dots,\}\subset{\Cal
C}_0(\Bbb R)_+$ is a sequence in the unit ball converging pointwise to $I$, then, for each $B\in\text{\got A}$,
$$\|[f_n(A),B]\|_{n\to\infty}\to0\ .$$
We will refer to $E$ as the {\it support of definition\/} of $A$ (or just support, if no confusion is possible).

If {\got A} has no unit, $A$ {\it is said to be affiliated to} {\got A} {\it if it is affiliated to  the multiplier
algebra\/} $M(\text{\got A})$.

A state $\omega$ (a representation $\pi$) of {\got A} is {\it in the support\/} of $A$ if $\tilde\omega(E)=1$ 
(resp. $\tilde\pi(E)=I$), where $\tilde\omega$, $\tilde\pi$ denote the normal extensions to {\got A}$^{**}$.

The normal extension of $f\to f(A)\in\text{\got A}$ to {\it bounded Borel functions\/} $f$ on $\Bbb R$ will be 
also denoted $f\in{\Cal B}(\Bbb R)\to f(A)\in\text{\got A}^{**}$. If $E(\lambda)\equiv\chi_{(-\infty,\lambda]}(A)$,
$\lambda\in\Bbb R$, $E$ is the strong limit of $E(\lambda)$ as $\lambda\to\infty$ and $\lambda\in\Bbb
R\to\tilde\pi(E(\lambda))$ is a {\it spectral family\/} for each $\pi$ in the support of $A$. For such $\pi$,
$\pi(A)$ will denote the selfadjoint operator with spectral resolution $\tilde\pi(E(\lambda))$.

We will say that the selfadjoint operator $A$ affiliated to {\got A} is
{\it central\/} if $f(A)$ is in the  centre
of {\got A} (or of $M(\text{\got A})$, if {\got A} has no unit) for each $f\in{\Cal C}_0(\Bbb R)$, i.e. if, for
each representation in the doman of $A$,
$$\pi(A)\;\;\eta\;\;\pi(\text{\got A})'\cap\pi(\text{\got A})''\ .$$
Since $\pi(A)$ is anyway affiliated to $\pi(\text{\got A})''$, 
$\pi(A)\;\;\eta\;\;\pi(\text{\got A})'$ would suffice.

A state $\omega\in{\Cal S}(\text{\got A})$ will be said to be in the {\it domain of\/} $A$ if it is in  the support
of $A$ and
$$\sup\{\omega(f(A));\ f\in{\Cal C}_0(\Bbb R)_+\ ,\quad f(\lambda)\leq\lambda^2\ ,\ \lambda\in\Bbb R\}\equiv
\omega(A^2)<\infty\ .$$

In this case we may and will write
$$\omega(A)=\int^{+\infty}_{-\infty}\lambda d\tilde\omega(E(\lambda))\ .$$
We can now state some easy facts.\medskip

I. A state $\omega\in{\Cal S}(\text{\got A})$ is in the support of definition of $A$ if and only  if if the same
applies to the GNS representation $\pi_\omega$.

Furthermore, $\omega$ is in the domain of $A$ if and only if $\xi_\omega$ is in the domain of  $\pi_\omega(A)$, in
which case we have
$$\omega(A)=(\xi_\omega,\pi_\omega(A)\xi_\omega)\ .$$
If $\omega$ is in the domain of $A$ we may define
$$(\Delta_\omega A)^2=\omega(A^2)-\omega(A)^2=\|(\pi_\omega(A)-\omega(A)I)\xi_\omega\|^2\ .$$
Of course we will say that $\omega$ is {\it definite\/} on $A$ if $\omega(A^2)=\omega(A)^2$, i.e.  if
$\pi_\omega(A)\xi_\omega=\omega(A)\xi_\omega$.

II. Let $\nu$ be a regular probability measure on ${\Cal S}(\text{\got A})$ with barycentre  $\omega\in{\Cal
S}(\text{\got A})$. Then $\omega$ is in {\it the support\/} of $A$ iff $\nu$ is carried by the states in the
support of $A$.

Moreover, $\omega$ is in the {\it domain\/} of $A$ if and only if $\nu$ is carried by the states  in the domain of
$A$ and $\varphi\to\varphi(A^2)$ is $\nu$--integrable. This implies that $\varphi\to\varphi(A)$ is $\nu$--square
integrable since $\varphi(A)^2\leq\varphi(A^2)$, hence $\varphi\to\varphi(A)$ is integrable and
$\varphi\to\Delta_\varphi A$ is square integrable.

If $\omega$ is also in the domain of another selfadjoint operator $B$ affiliated to {\got A},  we have as in
Proposition 3.3,
$$\Delta_\omega A\geq\int\Delta_\varphi Ad\nu(\phi)\ ,$$
$$\Delta_\omega A\Delta_\omega B\geq\int\Delta_\varphi A\Delta_\varphi Bd\nu(\varphi)\ .$$

III. If $\alpha$ is an automorphism of {\got A} and $A$ a selfadjoint operator affiliated to {\got A},  we can
define $\alpha(A)$ affiliated to {\got A} by the homomorphism
$$f\in{\Cal C}_0(\Bbb R)\to f(\alpha(A))\in M(\text{\got A})$$
s.t. $f(\alpha(A))=\tilde\alpha(f(A))$, where $\tilde\alpha\in\text{ Aut }M(\text{\got A})$ is the extension of
$\alpha$ defined by
$$\alpha(BC)=\tilde\alpha(B)\alpha(C)\ ,\qquad B\in M(\text{\got A})\ ,\qquad C\in\text{\got A}\ .$$

IV. If $A_1,\dots,A_n$ are selfadjoint operators affiliated to {\got A}, we say that they commute if  the
$f_i(A_i)$ commute with one another for each $f_i\in{\Cal C}_0(\Bbb R)$, $i=1,2,\dots,n$. In this case, the map
$f_1\otimes\dots\otimes f_n\in{\Cal C}_0(\Bbb R^n)\to f_1(A_1)\dots f_n(A_n)\in M(\text{\got A})$ extends to a
$*$--homomorphisms ${\Cal C}_0(\Bbb R^n)$ into $M(\text{\got A})$ whose support is called the {\it joint
spectrum\/} of $A_1,\dots,A_n$.

V. If $A$ is a selfadjoint operator affiliated to a $C^*$--algebra {\got A} and $\omega$ is a state on  {\got A} in
the domain of $A$, the state of {\got A} induced by the vector $\omega(A^2)^{-1/2}\pi_\omega(A)\xi_\omega$ in the
representation $\pi_\omega$ will be denoted by $\omega_A$. If $A$, $B$ are selfadjoint operators affiliated to
{\got A}, we will say that $\omega$ is in the domain of their commutator if $\omega_A$ is in the domain of $B$ and
$\omega_B$ in the domain of $A$, i.e. if $\xi_\omega$ is in the domain of $[\pi_\omega(A),\pi_\omega(B)]$. If $A$,
$B$, $C$ are selfadjoint operators affiliated to {\got A} with the same support $E$ such that, for each
representation $\pi$ in that support $(\tilde\pi(E)=I)$, the commutator $[\pi(A),\pi(B)]$ has closure $i\pi(C)$, we
will write
$$[A,B]^-=iC\ .$$

\beginsection{Appendix B}

I. {\it The manifold $\Sigma_+$ is homeomorphic to $TS^2$, the tangent bundle to the two--dimensional sphere\/}.

Let $\{\vec n,\vec v\}$ denote the generic point in $TS^2$, where $\vec n\in S^2$ is a unit vector  in $\Bbb R^3$
and $\vec v\in\Bbb R^3$ is s.t. $\vec v\cdot\vec n=0$.

For each $\sigma\in\Sigma_+$ let $\sigma=(\vec e,\vec m)$ be the parametrization by electric and  magnetic
components, so that $\vec e{\,}^2=\vec m^2$, $\vec e\cdot\vec m=1$. Hence $\vec e+\vec m\not=0$ and we can define
$$\vec n_\sigma\equiv\|\vec e+\vec m\|^{-1}(\vec e+\vec m)\ .$$
If $\sigma\in\Sigma_+$, $\sigma=(\vec e,\vec m)$ with $\vec e=\vec m$, let $h(\sigma)\in TS^2$ be defined by
$$h(\sigma)=\{\vec n_\sigma,\vec0\}=\{\vec e,\vec0\}=\{\vec m,\vec0\}\ .$$
If $\vec e\not=\vec m$, let $\vec u_\sigma=\|\vec e\times\vec m\|^{-1}\vec e\times\vec m$, and let  $L_\sigma$ be
the boost along $\vec u_\sigma$ with speed $\beta>0$ s.t.
$$\vec e{\,}^2=\vec m^2={1+\beta^2\over1-\beta^2}\ .$$
Then $L_\sigma$ takes $\sigma'=(\vec n_\sigma,\vec n_\sigma)\in\Sigma_+$ to $\sigma=(\vec e,\vec m)\in\Sigma_+$:
$$L_\sigma\sigma' L^T_\sigma=\sigma\ .\eqno(B.1)$$
Choose $\chi\in[0,+\infty)$ s.t. $\vec e^2=\vec m^2=\cosh4\chi$, i.e. $\gamma=(1-\beta^2)^{-1/2}=\cosh2\chi$.  We can
define a homeomorphism $h:\sigma\in\Sigma_+\to h(\sigma)\in TS^2$ by setting
$$h(\sigma)=\{\vec n_\sigma,\vec v_\sigma\}\ ;\qquad\vec v_\sigma=\chi\vec u_\sigma\ .\eqno(B.2)$$

II. {\it For each point $\vec n_\sigma\in S^2$ there is a Borel section for the action of $L^\uparrow_+$  on
$\Sigma_+$:
$$\sigma\in\Sigma_+\to\Lambda_\sigma\in L^\uparrow_+\ ,\eqno(B.3)$$
$$\Lambda_\sigma\sigma_0\Lambda^T_\sigma=\sigma\ ,\qquad\sigma\in\Sigma_+\ ,$$
where $\sigma_0=(\vec e_0,\vec e_0)$, $\vec e_0=(0,1,0)$, s.t.
$$\Lambda_\sigma=L_\sigma R_\sigma\ ;\quad L_\sigma\text{\it a boost, $R_\sigma$ a rotation,}\eqno(B.4)$$
$$\sigma\to L_\sigma\quad\text{\it is continuous on }\Sigma_+\ ,\eqno(B.5)$$
$$\sigma\in\Sigma_+\backslash\{\sigma/\vec n_\sigma=\vec n_0\}\to R_\sigma\text{ is continuous.}\
\eqno(B.6)$$}

We let $L_\sigma$ be defined as in I; it suffices to choose a continuous map $\vec n\in S^2\backslash \{\vec
n_0\}\to R(\vec n)$ from the 2--sphere minus one point to the rotation group such that $R(\vec n)\vec e_0=\vec n$.
We can take $R(\vec n_0)$ to be an arbitrary rotation taking $\vec e_0$ to $\vec n_0$ and set
$$R_\sigma=R(\vec n_\sigma)\ .$$
If $\vec n_0\neq-\vec e_0$, the map $R(\vec n)$ can be defined as the identity if $\vec n=\vec e_0$  or as the
rotation around $\vec e_0\times\vec n$ which takes $\vec e_0$ to $\vec n$ if $\vec n\not=\vec e_0$. If $\vec n_0$
is arbitrary, replacing $\vec e_0$ by $-\vec n_0$ in the above construction yields a map $R'(\vec n)$ and, for any
fixed choice of a rotation $R$ which takes $\vec e$ to $\vec n_0$, $R(\vec n)\equiv R'(\vec n)R$ meets the desired
requirements.

If $P$ denotes the space reflection, $\sigma_P=(-\vec e,\vec m)\in\Sigma_\mp$ if $\sigma=(\vec e, \vec
m)\in\Sigma_\pm$; we can define $\Lambda_\sigma\equiv PL_{\sigma_P}R_{\sigma_P}$, $\sigma\in\Sigma$, thus extending
our section to $\Sigma$ and letting it take values in the full Lorentz group.

Finally we point out yet another picture of $\Sigma$ which might turn out to be useful.  Associate to
$\sigma\in\Sigma$, $\sigma=(\vec e,\vec m)$, the vector $\vec u=\vec e+i\vec m\in\Bbb C^3$. Then
$v^2=\Sigma_jv^2_j=e^2-m^2+2i\vec e\cdot\vec m$, so that the image of $\Sigma$ is the manifold $\{v\in\Bbb
C^3/v^2=\pm2i\}$. The action of $L^\uparrow_+$ on $\Sigma$ corresponds to the action of ${\Cal O}(3,\Bbb C)$ on
that manifold, the space or time reflection to complex conjugation, and the Euclidean norm $\|\sigma\|$,
$\sigma\in\Sigma$ to the norm $\|v\|={1\over\sqrt2}\,(v,v)^{1/2}=\left({1\over2}\,\Sigma_j|v_j|^2\right)^{1/2}$.

\beginsection{Appendix C}

The goal of this appendix is to calculate the kernel ${\Cal D}_n$ and we begin by calculating the kernels 
describing the multiplication in ${\Cal E}$.

Let $f(q)=\int d^4k\hat f(k)e^{ikq}$. The function $f(x)=\int d^4k\hat f(k)e^{ikx}$ is called the symbol  of
$f(q)$. We formulate the multiplication in ${\Cal E}$ in terms of these symbols. The symbol of a product is
$$(f_1\dots f_n)(x)=\int d^4x_1\dots d^4x_nf_1(x_1)\dots f_n(x_n)C_n(x_1-x,\dots,x_n-x)\ ,\eqno(C.1)$$
where the distribution $C_n$ is given by
$$C_n(x_1,\dots,x_n)=(2\pi)^{-4}\int d^4k_1\dots d^4k_ne^{-i\Sigma k_jx_j+{i\over2}\,\Sigma_{j<l}
k_jQk_l}\ .\eqno(C.2)$$
The quadratic form in the exponential can be written in the form
$$\sum_{j<l}k_jQk_l={1\over2}\,(\underline k,B\otimes Q\underline k)\ ,\eqno(C.3)$$
where $\underline k=(k_1,\dots,k_n)\in\Bbb R^n\otimes\Bbb R^4$ and $B$ is an $n\times n$--matrix with entries 
$B_{jl}=1$, $j<l$, $B_{jl}=0$, $j=l$, $B_{jl}=-1$, $j<l$. If $n$ is even, the matrix $B$ has an inverse $B^{-1}$,
with entries $(B^{-1})_{jl}=(-1)^{j+l}B_{jl}$. Hence in this case we obtain
$$C_n(x_1,\dots,x_n)=\pi^{-2n}e^{-2i\Sigma_{j<l}(-1)^{j+l}x_jQ^{-l}x_l}\ ,\quad n\text{ even,}\eqno(C.4)$$
where we have used the fact that $B$ and $Q$ have determinant one.

For later convenience we introduce coordinates $y_j=\sum^j_{l=1}(-1)^lx_l$. In these coordinates the quadratic 
form in the exponential assumes the simple form
$$\sum_{j<l}(-1)^{j+l}x_jQ^{-1}x_l=\sum^{n-1}_{j=1}y_jQ^{-1}y_{j+1}\ .\eqno(C.5)$$
If $n$ is odd the matrix $B$ is not invertible. In this case we set $f_1,\dots,f_n=f_1,\dots,f_nf_{n+1}$ with 
$f_{n+1}=1$.

Since the symbol of the unit is the function $f_{n+1}(x)=1$ we obtain $C_n$ from $C_{n+1}$ by integrating over the
last variable. Hence
$$\eqalignno{
C_n(x_1,\dots,x_n)&=\int d^4x_{n+1}C_{n+1}(x_1,\dots,x_{n+1})\cr
&=\pi^{-2(n+1)}e^{-2i\Sigma^{n-1}_{j=1}y_jQ^{-1}y_{j+1}}\int d^4y_{n+1}e^{-2iy_nQ^{-1}y_{n+1}}\cr
&=\pi^{-2(n+1)}e^{-2i\Sigma^{n-1}_{j=1}y_jQ^{-1}y_{j+1}}(2\pi)^4\delta^{(4)}(2Q^{-1}y_n)\cr
&=\pi^{-2(n-1)}e^{-2i\Sigma^{n-2}_{j=1}y_jQ^{-1}y_{j+1}}\delta^{(4)}(y_n)\cr
&=C_{n-1}(x_1,\dots,x_{n-1})\delta^{(4)}\left(\sum^n_{j=1}(-1)^jx_j\right)\ ,\ n\text{ odd.}&(C.6)\cr}$$
If we replace $x_j$ by $x_j-x$ the coordinates $y_j$ with $j$ even remain unchanged whereas for $j$ odd $y_j$  is
replaced by $y_j+x$. We have
$$\sum^{2n}_{j=1}y_jQ^{-1}y_{j+1}=\sum^n_{j=1}y_{2j}Q^{-1}(y_{2j+1}-y_{2j-1})\ ,\eqno(C.7)$$
hence this expression will remain invariant.

Therefore we find
$$C_{2n}(x_1-x,\dots,x_{2n}-x)=\pi^{-4n}e^{-2i\Sigma^{2n-1}_{j=1}y_jQ^{-1}y_{j+1}}
e^{-2ixQ^{-1}y_{2n}}\eqno(C.8)$$
and
$$\eqalignno{
C_{2n+1}(x_1-x,\dots,x_{2n+1}-x)&=\pi^{-4n}e^{-2i\Sigma^{2n-1}_{j=1}y_jQ^{-1}y_{j+1}}
e^{-2ixQ^{-1}y_{2n}}\delta^{(4)}(y_{2n+1}+x)\cr
&=\pi^{-4}e^{-2i\Sigma^{2n}_{j=1}y_jQ^{-1}y_{j+1}}\delta^{(4)}(y_{2n+1}+x)\ .&(C.9)\cr}$$
The integral over a spatial hyperplane $x^0=t$ can now be easily performed. We obtain
$$\int_{x^0=t}d^3xC_{2n+1}(x_1-x,\dots,x_{2n+1}-x)=\pi^{-4n}e^{-2i\Sigma^{2n}_{j=1}y_jQ^{-1}y_{j+1}}
\delta(y^0_{2n+1}+t)\eqno(C.10)$$ in the odd case and
$$\eqalignno{
\int_{x^0=t}&d^3xC_{2n+2}(x_1-x,\dots,x_{2n+2}-x)\cr
&={1\over2\pi}\,\int d\eta\int d^4xe^{i\eta(x^0-t)}C_{2n+2}(x_1-x,\dots,x_{2n+2}-x)\cr
&={1\over2\pi}\,\int d\eta\int d^4x\pi^{-4(n+1)}e^{-2i\Sigma^{2n+1}_{j=1}y_jQ^{-1}y_{j+1}}
e^{-2ixQ^{-1}\left(y_{2n+2}-{1\over2}\,Q(\eta,\vec0)\right)}e^{-i\eta t}\cr
&={1\over2\pi}\,\int d\eta\pi^{-4}e^{-2i\Sigma^{2n}_{j=1}y_{j+1}}e^{-i(y^0_{2n+1}+t)}
\delta^{(4)}\left(y_{2n+2}-{1\over2}\,Q(\eta,\vec0)\right)&(C.11)\cr}$$
in the even case. Using (C.7) and the formulas
$$y_{2j}=\sum^j_{l=1}(x_{2l}-x_{2l-1})\ ,\quad y_{2j+1}-y_{2j-1}=x_{2j}-x_{2j+1}\ ,$$
$$y^0_{2n+1}+t=\sum^{2n+1}_{j=1}(-1)^j(x^0_j-t)\ ,\eqno(C.12)$$
we obtain the kernels
$${\Cal D}_n(x_1,\dots,x_n;t)=\int_{x^0=t}d^3xC_n(x_1-x,\dots,x_n-x)$$
in the form
$${\Cal D}_{2n+1}(x_1,\dots,x_{2n+1};t)=\pi^{-4n}e^{2iA_n(x_1-x_2,\dots,x_{2n}-x_{2n+1})}
\cdot\delta\left(\sum^{2n+1}_{j=1}(-1)^j(x^0_j-t)\right)\eqno(C.13)$$
and
$${\Cal D}_{2n+2}(x_1,\dots,x_{2n+2};t)=\pi^{-4n}e^{2iA_n(x_1-x_2,\dots,x_{2n}-x_{2n+1})}\cdot$$
$$\cdot{1\over2\pi}\,\int d\eta e^{i\eta\Sigma^{2n+1}_{j=1}(-1)^j(x^0_j-t)}
\delta^{(4)}\left(\sum^{n+1}_{j=1}(x_{2j}-x_{2j-1})-{1\over2}\,Q(\eta,\vec0)\right)\ ,
\eqno(C.14)$$
where $A_n$ is the quadratic form on $\Bbb R^{2n}\times\Bbb R^4$ given by
$$A_n(u_1,\dots,u_{2n})=\sum_{j\leq l\leq n}u_{2j-1}Q^{-1}u_{2l}\ .\eqno(C.15)$$\bigskip

\noindent{\it Acknowledgements\/}. It is our pleasure to thank D. Buchholz, A. Connes and  R. Haag for discussions.

We gratefully acknowledge the support given by the Graduiertenkolleg ``Theoretische Elementarteilchenphysik''  for
a visit of S.D. to Hamburg and by the CNR and the Dipartimento di Matematica, University of Rome I for visits of
K.F. to Rome thus making our collaboration possible.

S.D. takes pleasure in thanking Colin Sutherland for the warm hospitality and ideal working  conditions offered to
him at the Department of Pure Mathematics, University of New South Wales, in a stage of this research.\medskip

{\bf Note added in proof}: The model of QST proposed here is the simplest but not the unique  one implementing our
Uncertainty Relations (2.8), (2.9). In other models with this property the commutators of the q's are no longer
central, the associated algebras admit representations where the stronger uncertainty relations deduced in Section
2 are satisfied, namely $\inf(\Delta q_j$; $j=1,2,3)\cdot\sup(\Delta q_k,k=1,2,3)\gtrsim1$ holds too
in these representations, however, translation invariance at the
Planck scale is spontaneously broken [27].

We are grateful to A. Kempf for pointing out to us references [28,29].

\beginsection{References}

\item{1.} Amati, D., Ciafaloni, M., Veneziano, G.: Nucl. Phys. {\bf B347}, 551 (1990);  Amati, D.: On Spacetime at
Small Distances. In: Sakharov Memorial Lectures, Kladysh, L.V., Fainberg, V.Ya., eds., Nova Science P. Inc., 1992
\item{2.} Ellis, J., Mavromatos, N.E., Nanopoulos, D.V.: A Liouville String Approach to  Microscopic Time in
Cosmology. Preprint CERN--TH. 7000/93
\item{3.} Ashtekar, A.: Quantum Gravity: A Mathematical Physics Perspective. Preprint CGPG--93/12--2
\item{4.} Kempf, A.: Uncertainty Relations in Quantum Mechanics with Quantum Group Symmetry. Preprint DAMT/93--65
\item{5.} Woronowicz, S.L.: Compact Matrix Pseudogroups. Commun. Math. Phys. {\bf 111}, 613--665 (1987)
\item{6.} Wheeler, J.A.: Geometrodynamics and the Issue of the Final State. In: Relativity,  Groups and Topology.
De Witt, C., De Witt, B., (eds.) Gordon and Breach 1965. Hawking, S.W., Spacetime Foam. Nucl. Phys. {\bf B144},
349, (1978)
\item{7.} Madore, J.: Fuzzy Physics. Ann. Phys. {\bf 219}, 187--198 (1992)
\item{8.} Connes, A.: Non--Commutative Geometry. Academic Press, 1994
\item{9.} Connes, A., Lott, J.: Particle Models and Non Commutative Geometry. Nucl. Phys. B. (Proc. Suppl.) 
{\bf11B}, 19--47 (1990); Kastler, D.: A detailed account of Alain Connes' version of the standard model in
non--commutative geometry, I, II. Rev. Math. Phys. {\bf 5}, 477--532 (1993); III, Preprint CPT--92/P. 2814; IV
(with Th. Sch\"ucker), Preprint CPT--94/P.3092
\item{10.} Schrader, R.: The Maxwell Group and the Quantum Theory of Particles in Classical  Homogeneous Elecric
Fields. Forts. der Phys. {\bf 20}, 701--734 (1972)
\item{11.} Filk, T.: Field Theory on the Quantum Plane. Preprint, University of Freiburg, August 1990, THEP 90/12
\item{12.} Grosse, H., Madore, J.: A Noncommutative Version of the Schwinger Model. Phys. Lett.  {\bf B283}, 218
(1992)
\item{13.} Mack, G., Schomerus, V.: Models of Quantum Space Time: Quantum Field Planes. Preprint,  Harvard
University HUTMP 93--B335. Lukierski, J., Ruegg, H.: Quantum $\kappa$--Poincar\'e in any Dimension. Phys. Lett. B,
to appear
\item{14.} Doplicher, S., Fredenhagen, K., Roberts, J.E.: Spacetime Quantization Induced by  Classical Gravity.
Phys. Lett. B {\bf 331}, 33--44 (1994)
\item{15.} Hawking, S.W., Ellis, G.F.R.: The large scale structure of spacetime. Cambridge: Cambridge U.P., 1973
\item{16.} Knight, J.M.: Strict localization in Quantum Field Theory. J. Math. Phys. {\bf 2},  439--471 (1961);
Licht, A.L.: Strict localization. J. Math. Phys. {\bf 4}, 1443 (1963)
\item{17.} Haag, R.: Local Quantum Physics. Berlin, Heidelberg, New York: Springer TMP, 1993
\item{18.} Straumann, N.: General Relativity and Relativistic Astrophysics. Berlin--Heidel\-berg--New York: 
Springer, 1984
\item{19.} Woronowicz, S.L.: Unbounded elements affiliated with $C^*$--algebras and non--compact  Quantum Groups.
Commun. Math. Phys. {\bf 136}, 399--432 (1991). Woronowicz, S.L.: $C^*$--Algebras Generated by Unbounded Elements.
Preprint 1994; Baaj, S.: Multiplicateur non born\'e. Th\`ese, Universit\'e Paris VII 1980
\item{20.} Pedersen, G.K.: $C^*$--Algebras and their Automorphism Groups. New York: Acad. Press, 1979
\item{21.} Dixmier, J.: $C^*$--Algebras. Amsterdam: North Holland, 1980
\item{22.} Von Neumann, J.: \"Uber die Eindeutigkeit der Schr\"odingerschen Operatoren.  Math. Annalen {\bf 104},
570--578 (1931)
\item{23.} Filk, T.: Renormalizability of Field Theories on Quantum Spaces,  Preprint University of Freiburg, THEP
94/15, 1994
\item{24.} Didonato P. et al. eds.: Sympletic Geometry and Mathematical Physics. Actes du Colloque en Honneur de
J.M. Souriau, Basel, Boston: Birkh\"auser, 1991
\item{25.} Rieffel, M.A.: Deformation Quantization of Heisenberg Manifolds. Commun.  Math. Phys. {\bf 122},
531--562 (1989)
\item{26.} Landsman, N.P.: Strict Deformation Quantization of a Particle in External  Gravitational and Yang Mills
Fields. J. Geom. Phys. {\bf 12}, 93--132 (1993)
\item{27.} Doplicher, S., Fredenhagen, K.: in preparation
\item{28.} Maggiore, M.: Quantum Groups, Gravity, and the generalized uncertainty  principle, Phys. Rev. D{\bf49},
5182--5187 (1994); Phys. Lett. B {\bf 304}, 65--69 and {\bf 319}, 83--86 (1993)
\item{29.} Garay, L.J.: Quantum Gravity and minimum Length. Preprint Imperial/TP/93--94/20, gr--qc/9403008
\bye